\documentclass[aps,pre,superscriptaddress,showpacs,showkeys,nofootinbib]{revtex4}
\usepackage{amssymb,amsmath,amsfonts}
\usepackage{graphicx}
\usepackage{color}
\usepackage{url}
\usepackage{xfrac}
\usepackage[letterpaper,margin=1.0in]{geometry}

\newcommand{\calE}{{\mathcal E}}
\newcommand{\calO}{{\mathcal O}}

\newcommand{\taubar}{{\overline{\tau}}}
\newcommand{\zetabar}{{\overline{\zeta}}}

\begin{document}
\title{Oligarchy as a Phase Transition:  The effect of wealth-attained advantage in a Fokker-Planck description of asset exchange}
\thanks{\copyright 2016, all rights reserved}
\author{Bruce M. Boghosian}
\affiliation{Department of Mathematics, Tufts University, Medford, Massachusetts 02155, USA}
\author{Adrian Devitt-Lee}
\affiliation{Department of Mathematics, Tufts University, Medford, Massachusetts 02155, USA}
\author{Merek Johnson}
\affiliation{Department of Mathematics, Tufts University, Medford, Massachusetts 02155, USA}
\author{Jeremy A. Marcq}
\affiliation{Department of Mathematics, Tufts University, Medford, Massachusetts 02155, USA}
\author{Hongyan Wang}
\affiliation{Department of Mathematics, Tufts University, Medford, Massachusetts 02155, USA}
\date{\today}
\begin{abstract}
In earlier work~\cite{bib:Boghosian1,bib:Boghosian2}, we derived a nonlinear, integrodifferential Fokker-Planck equation for the ``Yard-Sale Model'' (YSM) of asset exchange.  In the absence of redistribution, we showed that the Gini coefficient $G$ is a Lyapunov functional for this model~\cite{bib:Boghosian3}, and that the time-asymptotic state of the model's wealth distribution has $G=1$, corresponding to complete inequality -- all of the wealth in the hands of a single agent.

When a simple one-parameter model of redistribution, based on the Ornstein-Uhlenbeck process, is introduced, we also showed that the model admits a steady state exhibiting some features in common with the celebrated Pareto Law of wealth distribution~\cite{bib:Boghosian1}.  In this work, we analyze the form of this steady-state distribution in much greater detail, using a combination of analytic and numerical techniques.  We find that, while Pareto's Law is approximately valid for low redistribution, it gives way to something more similar to Gibrat's Law when redistribution is higher.  Additionally, we prove in this work that, while this Pareto or Gibrat behavior persists over many orders of magnitude, it ultimately gives way to gaussian decay at extremely large wealth.

Following the work of Moukarzel et al.~\cite{bib:Moukarzel2007}, we then introduce a bias in favor of the wealthier agent in the YSM -- what we call Wealth-Attained Advantage (WAA) -- and show that this leads to the phenomenon of ``wealth condensation''~\cite{bib:BouchaudMezard2000} when the bias exceeds a certain critical value.  In the wealth-condensed state, a finite fraction of the total wealth of the population ``condenses'' to the wealthiest agent.  In this work, we examine this phenomenon in some detail, and derive the corresponding modification to the Fokker-Planck equation.  Earlier work~\cite{bib:Moukarzel2007} took the bias to be a discontinuous function of the wealth differential between the two transacting agents, and reported a first-order phase transition to absolute oligarchy.  By contrast, in this work we take the bias to be a continuous function of the wealth differential, and consequently we observe a second-order phase transition with a region of coexistence between the oligarch and a distribution of non-oligarchs.

We additionally show that the onset of wealth condensation has an abrupt reciprocal effect on the character of the non-oligarchical part of the distribution.  Specifically, we show that the above-mentioned gaussian decay at extremely large wealth is valid both above and below criticality, but degenerates to exponential decay precisely at criticality.
\end{abstract}
\pacs{89.65.Gh, 05.20.Dd}
\keywords{Fokker-Planck equation, Asset Exchange Model, Yard-Sale Model, H Theorem, Pareto distribution, Gibrat's Law, Lorenz curve, Gini coefficient, Lorenz-Pareto exponent, phase transitions, phase coexistence, wealth condensation}
\maketitle
\tableofcontents

\section{Introduction}
\label{sec:intro}
\subsection{Motivation and prior work}

Asset Exchange Models (AEMs) were first proposed by Angle~\cite{bib:Angle} in 1986 in the social sciences literature.  AEMs are collections of $N$ economic agents, each of which possesses some amount of wealth, and engages in pairwise transactions according to certain idealized rules.  These interactions are usually designed to conserve both the total number of agents $N$ and the total wealth $W$ for a closed economy, though it is also certainly possible to extend the model to account for production, consumption, immigration and emigration.  In the continuum limit, the wealth distribution can be described by an agent density function $P(w,t)$, the first two moments of which are the total number of agents $N = \int_0^\infty dw\; P(w,t)$ and the total wealth $W = \int_0^\infty dw\; P(w,t) w$.

In 1998, Ispolatov, Krapivsky and Redner~\cite{bib:IspolatovKrapivskyRedner} showed how to derive a Boltzmann equation for the time evolution of $P(w,t)$ for a particular AEM, in which the losing agent is selected with even odds and the amount lost is a fraction of the wealth of the losing agent.  This was the first application of the methods of modern mathematical physics to the analysis of AEMs.

A more economically realistic AEM was then proposed by Chakraborti~\cite{bib:Chakraborti2002} in 2002, and lucidly described by Hayes~\cite{bib:Hayes} shortly afterward who labelled it the ``Yard Sale Model'' (YSM).  In this model, the losing agent is still selected with even odds, but the amount lost is a fraction $\beta$ of the wealth of the {\it poorer} agent.  This assumption is perhaps more realistic because most economic agents engage in transactions for which the amount at stake is strictly less than their own total wealth.

In 2007, Moukarzel et al.~\cite{bib:Moukarzel2007} modified the YSM by biasing the choice of losing agent so as to confer an advantage to the wealthier of the two transacting agents.  In the present paper, we refer to such a modification as Wealth-Attained Advantage (WAA).  Moukarzel et al.\ analyzed the steady-state master equation for the YSM with WAA, and demonstrated that it exhibits a first-order phase transition to a fully {\it wealth-condensed} state.  Wealth condensation, which was first reported by Bouchaud and M\'{e}zard in 2000 and later studied by Burda et al.~\cite{bib:BurdaJohnstonEtAl2002}, is characterized by the concentration of a finite fraction of wealth in the hands of a single agent, and it is thought to provide a statistical mechanical explanation of the phenomenon of oligarchy.

In 2014 Boghosian~\cite{bib:Boghosian1} derived the Boltzmann equation for the YSM, and showed that it reduces to a nonlinear, integrodifferential Fokker-Planck equation in the limit of small $\beta$, which he termed the {\it small-transaction limit}.  The derivation is similar to that of the nonlinear, integrodifferential Fokker-Planck collision operator used in plasma physics from the Boltzmann equation in the weak-collision limit~\cite{bib:RMJ}.  This Fokker-Planck description is significant for its {\it universality}:  Though different posited distributions for $\beta$ would result in different Boltzmann equations, the Fokker-Planck equation obtained from all of them in the small-transaction limit is universal in form.

Here an analogy may be made with kinetic theory:  While different dilute gases may exhibit very different collision dynamics at the molecular level, and therefore be described by very different Boltzmann equations, the Chapman-Enskog analysis teaches us that they all reduce to the Navier-Stokes equations in the limit of small Knudsen number.  The universality of the Fokker-Planck description for the macroscopic description of wealth distributions~\cite{bib:Boghosian1} is likewise analogous to that of the Navier-Stokes equation for the macroscopic description of dilute gases, and the small transaction limit plays the role of small Knudsen number in this metaphor.

Later in 2014, Boghosian~\cite{bib:Boghosian2} showed that this universal Fokker-Planck equation could be derived directly from the underlying stochastic process, without the intermediary of the Boltzmann equation.  A number of numerical studies~\cite{bib:Chakraborti2002,bib:Moukarzel2007,bib:Boghosian1,bib:Boghosian2} have presented numerical evidence that wealth concentrates without bound in the YSM, unless it is supplemented with some model for redistribution.  This latter point was then proven rigorously by Boghosian, Johnson and Marcq in 2015~\cite{bib:Boghosian3}, who showed that the Gini coefficient is a Lyapunov functional of both the Boltzmann and the universal Fokker-Planck equations for the YSM.  This work also demonstrated that the time-asymptotic value of the Gini coefficient of the non-redistributive YSM is unity, corresponding to absolute oligarchy.

At first glance, the instability of the YSM without redistribution seems counter-intuitive since the losing agent in any transaction is selected with even odds.  The fact that the amount at stake in any transaction is always a smaller fraction of the wealth of the poorer agent than of the richer agent, however, means that the latter is able to withstand a longer string of losses, and this ultimately breaks the symmetry.  It may be noted in passing that this is consistent with Keynesian economic theories which suggest that market economies are inherently unstable without some kind of government intervention -- e.g., redistribution.

When the YSM is stabilized by a one-parameter model of redistribution, based on the Ornstein-Uhlenbeck process~\cite{bib:OU}, numerical evidence has been presented~\cite{bib:Boghosian1,bib:Boghosian2} indicating that this redistributive YSM yields a steady-state that shares many features with the celebrated Pareto distribution~\cite{bib:Pareto}, especially at low values of the redistribution parameter.  In particular, it exhibits a lower cutoff, and an approximate power-law decay over several orders of magnitude.

\subsection{Contributions of this study}

In its simplest form, the Fokker-Planck equation for the redistributive YSM requires only two free parameters.  One is the number of transactions per unit time, and the other is the redistribution per unit time.  In steady state, only the ratio of these, namely the redistribution per transaction, matters.  We begin the present work by examining this steady state distribution and the Gini coefficient obtained for different values of this ratio, using a variety of analytic and numerical methods, and discussing the implications of these findings for real economies.  In this work we show that, while the Pareto law is valid for low redistribution, it gives way to something more similar to Gibrat's Law when redistribution is higher.  Additionally, we show that, while this Pareto or Gibrat behavior may persist over many orders of magnitude, it ultimately gives way to gaussian decay at extremely large wealth.

Next, we introduce WAA into the model and derive the correction to the Fokker-Planck equation corresponding to this modification.  Our model for WAA differs from that used by Moukarzel et al.~\cite{bib:Moukarzel2007}, in a way that has important physical consequences, and therefore warrants explanation.  The WAA model introduced in \cite{bib:Moukarzel2007} gives a winning probability, of fixed value $\sfrac{1}{2}+\epsilon$, to the wealthier agent, and its complement to the less wealthy agent, irrespective of the magnitude of the wealth difference.  The model introduced in this paper gives a winning probability to each agent of the form $\sfrac{1}{2}+\epsilon(w_> - w_<)$, where $w_>$ is the wealth of the wealthier agent, $w_<$ is that of the less wealthy agent.  That is, our bias is a smooth function of the wealth difference, vanishing when the two transacting agents have equal wealth.

As a consequence of the above-described smooth dependence of bias on wealth difference in our WAA model, the wealth condensation that we observe is a phase transition of one higher order than that observed in earlier work~\cite{bib:Moukarzel2007}.  If we think of the order parameter as either the fraction of wealth held by the wealthiest agent or the Gini coefficient, the model introduced by Moukarzel~\cite{bib:Moukarzel2007} exhibits a first-order phase transition at criticality to order parameter unity, and thus to complete oligarchy.  The model introduced in this paper, by contrast, exhibits a slope discontinuity in the order parameter at criticality, indicative of a second-order phase transition, above which the wealth-condensed phase is characterized by {\it coexistence} between the oligarch and the distribution of non-oligarchs.

In this work we also demonstrate that the onset of wealth condensation in the YSM has a reciprocal impact on the classical (non-oligarchical) portion of the wealth distribution.  In particular, the above-mentioned gaussian decay at extremely large wealth degenerates to exponential decay precisely at criticality.  Mathematically, the classical part of the wealth distribution decays like $\exp(-|a|w^2-bw)$, where the quantity $a$ passes through zero at the onset of wealth condensation.

\subsection{Structure of this paper}

In Sec.~\ref{sec:nonredist}, we review the derivation of the nonlinear, integrodifferential Fokker-Planck equation for the version of the YSM without redistribution, mostly in order to introduce notation, including the singular distribution, $\Xi(w)$, which is further described in Appendix~\ref{sec:Xi}.  This distribution, which is present in the time-asymptotic limit of the simplest version of the model, can appear at finite time when the level of WAA exceeds its critical value.

In Sec.~\ref{sec:redist}, we review the addition of redistribution to the model, so that we can (i) show how the Fokker-Planck equation can be rescaled to a simple canonical form, and (ii) present much more accurate numerical simulations than have been presented in our earlier work on the YSM -- e.g., in \cite{bib:Boghosian1,bib:Boghosian2}.  The numerical methods used are presented in Appendix~\ref{sec:numerical}.  We study the asymptotic behavior of the solution for small, intermediate and large $w$.  We also introduce the Gini coefficient as an order parameter, and investigate its dependence on the redistribution parameter.

Finally, in Sec.~\ref{sec:waa} we introduce WAA to the model, in the fashion described above.  We generalize the above-mentioned canonical form for this case, and we also generalize the analysis of the solution for small, intermediate and large $w$.  Both the analysis for intermediate $w$ and that for large $w$ yield evidence of wealth condensation when the WAA parameter exceeds the redistribution parameter.  This is demonstrated numerically for intermediate $w$ and analytically for large $w$, with the latter analysis appearing in Appendix~\ref{sec:LargeW}.  It is demonstrated that both the fraction of wealth held by the wealthiest agent and the Gini coefficient exhibit slope discontinuities at criticality, indicative of a second-order phase transition.

\section{Fokker-Planck equation for the non-redistributional Yard-Sale Model}
\label{sec:nonredist}
\subsection{The agent density function and the Pareto potentials}

In the simplest version of the YSM, agents have only a single attribute, namely their wealth $w$, which is assumed to be positive.  At the micro level, the economy evolves in time by sequential pairwise transactions between these agents.  At the macro level, in the continuum limit, we represent the wealth distribution by the agent density function, $P(w,t)$, which is non-negative and defined so that $\int_a^b dw\; P(w,t)$ is the number of agents with wealth between $a$ and $b$ at time $t$.  The goal is to determine how the dynamics of transacting agents at the micro level give rise to a dynamical equation for $P(w,t)$ at the macro level.  As noted in the Introduction, the first two moments of the agent density function are the total number of agents
\begin{equation}
N = \int_0^\infty dw\; P(w,t)\end{equation}
\label{eq:N}
and the total wealth
\begin{equation}
W = \int_0^\infty dw\; P(w,t) w.
\label{eq:W}
\end{equation}
Higher moments may or may not exist, depending on the asymptotics of $P(w,t)$.

The first systematic studies of the form of $P(w,t)$ for real economies were made by Pareto~\cite{bib:Pareto}, who tabulated the fraction of agents with wealth greater than $w$ as a function of $w$.  Using our notation, this fraction is given by
\begin{equation}
A(w,t) := \frac{1}{N}\int_w^\infty dx\; P(x,t).\;\;\;\;\;
\label{eq:A}
\end{equation}
Pareto asserted that $A(w,t)$ is approximately one for $w$ below a lower cutoff, and that it decays like the power law $w^{-\alpha}$ at large values of $w$.  The exponent $\alpha$ is called the Lorenz-Pareto exponent.  We shall compare results from the YSM with Pareto's observations later in this paper.

In what follows, it will also be useful to define the following incomplete moments of $P$,
\begin{eqnarray}
F(w,t) &:=& \frac{1}{N}\int_0^w dx\; P(x,t)\label{eq:F}\\
L(w,t) &:=& \frac{1}{W}\int_0^w dx\; P(x,t) x\label{eq:L}\\
B(w,t) &:=& \frac{1}{N}\int_0^w dx\; P(x,t) \frac{x^2}{2}.\label{eq:B}
\end{eqnarray}
Note that $A(w,t)+F(w,t)=1$.  The non-negativity of $P$ implies that $A$ is non-increasing, that $F$ and $L$ are non-decreasing, and that the range of $A$, $F$ and $L$ is $[0,1]$.  We shall refer to $A$, $F$, $L$ and $B$ collectively as the {\it Pareto potentials}.

\subsection{Transactional exchange}

At each step of the micro evolution of the YSM, a pair of agents is randomly selected to engage in a transaction.  To be concrete, we suppose that an agent with wealth $\overline{w}$ enters a transaction with another agent with wealth $\overline{x}$.  If these agents exchange items of equal value, no wealth changes hands.  If, however, one of them makes a ``mistake,'' due to a variety of processes -- such as asymmetric information, irrational expectations, behavioral economic issues, or a perverse utility function -- then a certain amount of wealth $\Delta w$ will change hands.  Instead of trying to develop reductionist models for these various processes, the YSM treats them stochastically, by asserting that $\Delta w$ is a constant $\beta$ times the smaller of the wealths of the two agents, and that the direction of wealth transfer is determined by the flip of a fair coin.

Mathematically, we suppose that the final wealth of the first agent is given by
\begin{equation}
w = \overline{w} + \Delta w,
\label{eq:transform}
\end{equation}
where
\begin{equation}
\Delta w = \sqrt{\gamma\Delta t}\,\min(\overline{w},\overline{x})\eta.
\label{eq:deltaW}
\end{equation}
Here we have taken $\beta=\sqrt{\gamma\Delta t}$, where $\Delta t$ is the characteristic time associated with a transaction, for reasons which will become clear shortly~\footnote{Throughout this work, we take the approach of introducing parameters that will simplify the resulting Fokker-Planck equation, even if they do not seem the most natural in the microscopic description of the random process.}.  The coin flip is modeled by the stochastic variable $\eta\in\{-1,+1\}$.

\subsection{Derivation of the Fokker-Planck equation for the non-redistributional YSM}

From Eqs.~(\ref{eq:transform}) and (\ref{eq:deltaW}), we see that the post-transaction wealth of the first agent depends on two stochastic variables, namely $\eta$ and $\overline{x}$.  Going forward, we are going to need to take averages over these variables.  We denote expectation values over functions of $\eta$ by $E$, and in particular we demand that
\begin{equation}
E[\eta]=0
\label{eq:Eeta}
\end{equation}
so that the coin flip is fair, and
\begin{equation}
E[\eta^2]=1
\end{equation}
for normalization.

The expectation value of a function $f(\eta,\overline{x})$ that depends on both $\eta$ and $\overline{x}$ at time $t$ is then denoted by
\begin{equation}
\calE[f] = \frac{1}{N}\int_0^\infty d\overline{x}\; P(\overline{x},t) E[f(\eta,\overline{x})].
\label{eq:average}
\end{equation}
Here we have assumed that the wealth $\overline{x}$ of the other agent is distributed according to the (unknown) probability density function $P/N$.

To derive a Fokker-Planck equation from a random process, we need to compute the {\it drift coefficient},
\begin{equation}
\sigma = \lim_{\Delta t\rightarrow 0}\calE\left[\frac{\Delta w}{\Delta t}\right] = 0,
\label{eq:sigma1}
\end{equation}
and the {\it diffusivity},
\begin{equation}
D = \lim_{\Delta t\rightarrow 0}\calE\left[\frac{\left(\Delta w\right)^2}{\Delta t}\right]
=
\frac{\gamma}{N}\int_0^\infty dx\; P(x,t)\left[\min(w,x)\right]^2
=
2\gamma\left[B(w,t) + \frac{w^2}{2}A(w,t)\right],
\label{eq:D1}
\end{equation}
where we have used Eqs.~(\ref{eq:deltaW}) and (\ref{eq:average}).  The Fokker-Planck equation describing a process with drift coefficient $\sigma$ and diffusivity $D$ is then
\begin{equation}
\frac{\partial P}{\partial t}
=
-\frac{\partial}{\partial w}\left(\sigma P\right) + \frac{1}{2}\frac{\partial^2}{\partial w^2}\left(D P\right),
\label{eq:fp}
\end{equation}
which in this case becomes the nonlinear, integrodifferential equation
\begin{equation}
\boxed{
\frac{\partial P}{\partial t}
=
\frac{\partial^2}{\partial w^2}\left[\gamma\left(B + \frac{w^2}{2}A\right) P\right],
}
\label{eq:fp1}
\end{equation}
where we have omitted functional dependences for ease of notation.  We adopt Eq.~(\ref{eq:fp1}) as the fundamental equation of motion describing the dynamics of the agent density function $P$ in a non-redistributive economy, in which the direction of transactional wealth transfer is determined by the toss of a fair coin.

\subsection{Conservation laws}

Later in this paper, we will show~\footnote{More precisely, in this paper we demonstrate this only in steady state, and leave the demonstration for the time-dependent case as an open problem.} that the quantity in square brackets in Eq.~(\ref{eq:fp1}) and its derivative with respect to $w$  both vanish faster than any polynomial in $w$ as $w\rightarrow 0$ and as $w\rightarrow\infty$.  From this observation, we can confirm that the model conserves agents and total wealth as follows,
\begin{eqnarray}
\frac{dN}{dt}
&=& \int_0^\infty dw\; \frac{\partial P}{\partial t}
= \int_0^\infty dw\; \frac{\partial^2}{\partial w^2}\left[\gamma\left(B + \frac{w^2}{2}A\right) P\right]
= \left.\frac{\partial}{\partial w}\left[\gamma\left(B + \frac{w^2}{2}A\right) P\right]\right|_0^\infty
= 0\\
\noalign{\noindent \mbox{and}}
\frac{dW}{dt}
&=& \int_0^\infty dw\; \frac{\partial P}{\partial t}w
= \int_0^\infty dw\; \frac{\partial^2}{\partial w^2}\left[\gamma\left(B + \frac{w^2}{2}A\right) P\right]w
= -\left.\left[\gamma\left(B + \frac{w^2}{2}A\right) P\right]\right|_0^\infty
= 0,
\end{eqnarray}
where we integrated by parts in the second instance.  Thus, as noted earlier, the basic transactional YSM describes the dynamics of $P$ in a closed economy, with no births, deaths, immigration, emigration, production or consumption.

\subsection{Lyapunov functional}

\begin{figure}
\begin{center}
\includegraphics[bbllx=0,bblly=0,bburx=360,bbury=376,width=.25\textwidth]{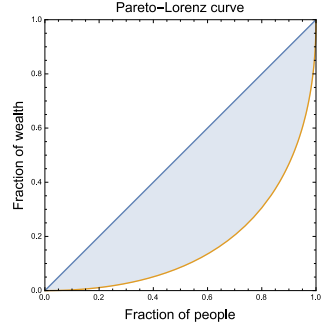}
\end{center}
\caption{{\bf The Lorenz curve} shows the fraction of the wealth held by agents with wealth less than $w$ versus the fraction of agents with wealth less than $w$.  The Gini coefficient is the ratio of the the shaded area to the area of the triangle below the diagonal.}
\label{fig:Pareto-Lorenz}
\end{figure}

The Gini coefficient is a well known measure of inequality.  Its minimum value of zero is achieved when all agents have exactly the same amount of wealth, and its maximum value of one is approached when all the wealth is in the hands of a vanishingly small number of people.  To define it geometrically, we begin with a parametric plot of $L(w,t)$ versus $F(w,t)$, for $0\leq w < \infty$.  This plot, illustrated in Fig.~\ref{fig:Pareto-Lorenz}, is called the {\it Lorenz curve}.  It can be thought of as the fraction of the wealth held by agents with wealth less than $w$ versus the fraction of agents with wealth less than $w$.  If all agents had exactly the same wealth, the Lorenz curve would coincide with the diagonal.  In practice, it can be shown that it always lies below the diagonal and is concave up (see, e.g., \cite{bib:Boghosian3}).  The Gini coefficient $G(t)$ is then defined as twice the area between the actual curve and the diagonal; this area is shaded in Fig.~\ref{fig:Pareto-Lorenz}.

It is straightforward to show (see, e.g., \cite{bib:Boghosian3}) that the Gini coefficient can be written as a functional of the agent density function as follows,
\begin{eqnarray}
G(t) &=& 1 - \frac{2}{W}\int_0^\infty dw\; P(w,t)A(w,t) w,
\label{eq:gini1}\\
\noalign{\noindent \mbox{or alternatively as}}
G(t) &=& 1 - \frac{2}{N}\int_0^\infty dw\; P(w,t)L(w,t),
\label{eq:gini2}
\end{eqnarray}
where $A(w,t)$ and $L(w,t)$ are the Pareto potentials defined in Eqs.~(\ref{eq:A}) and (\ref{eq:L}), respectively.  In recent work~\cite{bib:Boghosian3}, it has been demonstrated that $G$ is a Lyapunov functional of Eq.~(\ref{eq:fp1}),
\begin{equation}
\frac{dG}{dt}\geq 0,
\end{equation}
and that
\begin{equation}
\lim_{t\rightarrow\infty}  G(t) = 1,
\end{equation}
both for the Boltzmann and Fokker-Planck equations for a purely transactional economy (without redistribution).  Consequently, Eq.~(\ref{eq:fp1}) can only increase inequality so that the distribution $P(w,t)$ becomes singular in nature in the time-asymptotic limit.  In the following subsection, we characterize this singular limit more carefully.

\subsection{Time-asymptotic limit}
\label{ssec:Xi}

If our initial conditions have $N$ agents and total wealth $W$, it is worth investigating the nature of the singular distribution $\lim_{t\rightarrow\infty}P(w,t)$.  If the agents were discrete, as would be the case in a real economy, or might be the case in a Monte Carlo representation of the wealth distribution, then in the time-asymptotic limit one agent would end up with all the wealth $W$, while the other $N-1$ agents would have none at all; that is,
\begin{equation}
\lim_{t\rightarrow\infty} P(w,t) = (N-1)\delta(w) + \delta(w - W),
\end{equation}
where $\delta$ denotes the Dirac delta distribution.

Because we are using a continuum-valued distribution $P(w,t)$, however, $N$ is not restricted to the integers.  So there is nothing to prevent ``half an agent'' from taking wealth $2W$ while the other $N-\sfrac{1}{2}$ agents have zero wealth, or a ``third of an agent'' from taking wealth $3W$ while the other $N-\sfrac{1}{3}$ agents have zero wealth, etc.  Continuing along these lines, let's suppose that a fraction $\epsilon$ of an agent takes wealth $W/\epsilon$, while the other $N-\epsilon$ agents have wealth zero, so in the continuum limit we have the singular solution
\begin{eqnarray}
\lim_{t\rightarrow\infty} P(w,t)
&=&
\lim_{\epsilon\rightarrow 0}
\left[
(N-\epsilon)\delta(w) +
\epsilon\delta\left(w - \frac{W}{\epsilon}\right)
\right]\nonumber\\
&=&
N\delta(w) +
W \lim_{\epsilon\rightarrow 0}\epsilon\delta\left(w - \frac{1}{\epsilon}\right).
\end{eqnarray}
This may be written more succinctly as
\begin{equation}
\lim_{t\rightarrow\infty} P(w,t) = N\delta(w) + W\Xi(w),
\label{eq:distSol}
\end{equation}
where we have defined the distribution
\begin{equation}
\Xi(w) := \lim_{\epsilon\rightarrow 0}\epsilon\;\delta\left(w - \frac{1}{\epsilon}\right).
\label{eq:XiDef}
\end{equation}

Unfortunately, the limit that appears in the definition in Eq.~(\ref{eq:XiDef}) is ambiguous.  If we were to multiply both sides of Eq.~(\ref{eq:XiDef}) by an arbitrary test function $f(w)$ and integrate over $w$, it is not the least bit clear whether the integral can be exchanged with the limit.  To be specific, we would like to write
\begin{equation}
\int_0^\infty dw\; \Xi(w)f(w) =
\lim_{\epsilon\rightarrow 0}\epsilon\int dw\;\delta\left(w - \frac{1}{\epsilon}\right)f(w) =
\lim_{\epsilon\rightarrow 0}\epsilon f\left(\frac{1}{\epsilon}\right) =
\lim_{w\rightarrow\infty} \frac{f(w)}{w}.
\label{eq:XiDef2}
\end{equation}
One must be suspicious of such reasoning, however, because of the cavalier interchange of the order of the integral and the limit.  Moreover, it is evident that the limit in the rightmost expression in Eq.~(\ref{eq:XiDef2}) will not exist if $f$ grows faster than linearly as $w\rightarrow\infty$.

Functional analysis addresses this problem by encouraging us to think of $\Xi$ as a linear functional, rather than as a function, and it also encourages us to think carefully about the function space of allowed test functions $f$.  We therefore {\it define} the singular distribution $\Xi$ by the prescription
\begin{equation}
\int dw\; \Xi(w) := \lim_{w\rightarrow\infty} \frac{f(w)}{w},
\end{equation}
for the space of test functions $f$ for which the limit exists.  In particular, the first two moments of $\Xi$ are given by
\begin{eqnarray}
\int_0^\infty dw\; \Xi(w) &=& \lim_{w\rightarrow\infty} \frac{1}{w} = 0,\\
\int_0^\infty dw\; \Xi(w) w &=& \lim_{w\rightarrow\infty} \frac{w}{w} = 1,
\end{eqnarray}
demonstrating that $\Xi$ has a zeroth moment of zero, and a first moment of one.  From this perspective, higher moments of $\Xi$ do not even make sense because $w^n$ is not in the space of test functions unless $n\leq 1$.  A more rigorous and complete characterization of $\Xi$, using Sobolev-Schwarz distribution theory, was given in Appendix C of \cite{bib:Boghosian1}, where the singular distribution $\Xi$ was first presented.

From the above definition of $\Xi$, it follows that the distributional solution given in Eq.~(\ref{eq:distSol}) has the correct moments.  Intuititively, if $\delta$ represents one unit of agent corresponding to zero wealth, then $\Xi$ represents one unit of wealth corresponding to zero (infinitely wealthy) agents.  The fact that the Gini coefficient is a Lyapunov functional of Eq.~(\ref{eq:fp1}) indicates that Eq.~(\ref{eq:distSol}) is the time-asymptotic limit of the solution to Eq.~(\ref{eq:fp1}) for arbitrary initial conditions.

In the above presentation, we have defined the singular distribution $\Xi$ as a linear functional, which is also the proper way to think of the Dirac delta, but there is a more pedestrian way to think of these objects.  Just as $\delta$ can be understood as the limit of a sequence of functions, so can $\Xi$.  This approach is presented in Appendix~\ref{sec:Xi}.  

Enormous wealth in the hands of a vanishingly small fraction of the population is the signature of oligarchy.  Going forward, a distributional solution of the form
\begin{equation}
P(w,t) = p(w,t) + W_\Xi(t)\Xi(w),
\label{eq:oligarchicalSol}
\end{equation}
where $p$ has zeroth moment $N$ and first moment $W-W_\Xi(t)$, will be called an {\it oligarchical solution} whenever $W_\Xi(t)>0$.  When $W_\Xi(t)=W$, it will be said to be an {\it absolute oligarchy}.  The above arguments have established that the time-asymptotic state of the non-redistributive YSM is an absolute oligarchy.

\section{Redistribution in the Yard-Sale Model}
\label{sec:redist}
\subsection{Derivation of the Fokker-Planck equation for the redistributional YSM}

Absolute oligarchy is not observed in the real world.  Very few people have zero wealth, and indeed the number of people who live in extreme poverty seems to have decreased significantly over the past few decades~\cite{bib:ExtremePoverty}.  In earlier work~\cite{bib:Boghosian1}, we introduced a simple model of redistribution to stabilize the YSM and give rise to non-trivial steady-state solutions.  We review that redistribution model here and present some new results about its steady-state solutions.

Our basic model of redistribution supposes that we impose a ``wealth tax'' on each agent, with tax rate per unit time $\tau$, and parcel out the tax thereby collected in $N$ equal shares to the $N$ agents.  The tax collected from an agent with wealth $\overline{w}$ in one transaction time is then $(\tau\Delta t) \overline{w}$.  The total tax collected from everybody in one transaction time is therefore $(\tau\Delta t) W$, and a fraction $1/N$ of this is returned to the agent with wealth $\overline{w}$.  The net gain experienced by the agent with wealth $\overline{w}$ in a single transaction is then $(\tau\Delta t)(W/N - \overline{w})$.  This will be positive for agents with wealth less than the average value of $W/N$, and negative otherwise.

From the above accounting, we see that the effect of redistribution is to modify Eq.~(\ref{eq:deltaW}) as follows
\begin{equation}
\Delta w = (\tau\Delta t)(\frac{W}{N} - \overline{w}) + \sqrt{\gamma\Delta t}\,\min(\overline{w},\overline{z})\eta.
\label{eq:deltaWRedist}
\end{equation}
This gives rise to a nonzero drift term
\begin{equation}
\sigma = \calE\left[\frac{\Delta w}{\Delta t}\right] = \tau\left(\frac{W}{N} - w\right),
\label{eq:sigma2}
\end{equation}
while the diffusivity of Eq.~(\ref{eq:D1}) is unchanged.  The nonlinear, integrodifferential Fokker-Planck equation for the redistributive YSM is therefore
\begin{equation}
\boxed{
\frac{\partial P}{\partial t}
=
-\frac{\partial}{\partial w}\left[\tau\left(\frac{W}{N} - w\right)P\right] +
\frac{\partial^2}{\partial w^2}\left[\gamma\left(B + \frac{w^2}{2}A\right) P\right].
}
\label{eq:fp2}
\end{equation}
The redistributive term is similar in form to that in the famous Ornstein-Uhlenbeck model~\cite{bib:OU}, where it plays the role of stabilizing the PDF of the Wiener process.

That Eq.~(\ref{eq:fp2}) still conserves agents is manifest.  That it still conserves wealth follows from an integration by parts and the easily verified identity
\begin{equation}
0 = \int_0^\infty dw\; \tau\left(\frac{W}{N} - w\right)P(w,t).
\label{eq:ConsId1}
\end{equation}
The surface term in the integration by parts vanishes from the fact that $P$ decays to zero faster than any polynomial in $w$, both as $w\rightarrow 0$ and as $w\rightarrow\infty$.

\subsection{Scaling and canonical form}
\label{ssec:scaling}

As things stand, Eq.~(\ref{eq:fp2}) appears to have three parameters, namely $\gamma$, $\tau$ and $W/N$.  In this subsection, we show how to scale the equation to reduce the parametric study required to a single parameter.  We could accomplish this in a single transformation, involving both the dependent and independent variables, but it will seem more motivated if we do it in steps.

We first note that the only quantity in the equation that sets a scale for wealth is the average wealth, $W/N$.  It would therefore make sense to measure $w$ in units of $W/N$, so we define the dimensionless variable,
\begin{equation}
\overline{w} = \frac{w}{W/N}.
\label{eq:trans0}
\end{equation}
This induces a transformation in the agent density function.  The new version, $\overline{P}(\overline{w},t)$, must satisfy
\begin{equation}
\int_0^{w} dx\; P(x,t)
=
\int_0^{\overline{w}} d\overline{x}\; \overline{P}(\overline{x},t).
\end{equation}
Differentiating both sides of the above equation with respect to $w$ yields
\begin{equation}
P(w,t) = \frac{d\overline{w}}{dw}\overline{P}(\overline{w},t) = \frac{N}{W}\overline{P}(\overline{w},t).
\label{eq:trans1}
\end{equation}

In terms of the new agent density function, the Pareto potentials may be written
\begin{eqnarray}
A(w,t)
&=&
\frac{1}{N}\int_w^{\infty} dx\; P(x,t)
=
\overline{A}(\overline{w},t)\\
\noalign{\noindent \mbox{and}}
B(w,t)
&=&
\frac{1}{N}\int_0^w dx\; P(x,t) x^2
=
\frac{W^2}{N^2}
\overline{B}(\overline{w},t),
\end{eqnarray}
where we have defined
\begin{eqnarray}
\overline{A}(\overline{w},t)
& := &
\frac{1}{N}\int_{\overline{w}}^{\infty} d\overline{x}\; \overline{P}(\overline{x},t)\\
\noalign{\noindent \mbox{and}}
\overline{B}(\overline{w},t)
& := &
\frac{1}{N}\int_{\overline{w}}^{\infty} d\overline{x}\; \overline{P}(\overline{x},t) \frac{{\overline{x}}^2}{2}.
\end{eqnarray}
Rewriting Eq.~(\ref{eq:fp2}) with independent variable $\overline{w}$, we find
\begin{equation}
\frac{\partial \overline{P}}{\partial t}
=
-\frac{\partial}{\partial \overline{w}}\left[\tau\left(1 - \overline{w}\right)\overline{P}\right] +
\frac{\partial^2}{\partial {\overline{w}}^2}\left[\gamma\left(\overline{B} + \frac{{\overline{w}}^2}{2}\overline{A}\right) \overline{P}\right].
\label{eq:fp2a}
\end{equation}
In some sense, Eq.~(\ref{eq:fp2a}) is equivalent to solving the problem for $W/N=1$.  Once that solution for $\overline{P}$ is known, Eqs.~(\ref{eq:trans0}) and (\ref{eq:trans1}) tell us how to recover the agent density function $P$ for arbitrary $W/N$,
\begin{equation}
P(w,t) = \frac{N}{W}\overline{P}\left(\frac{w}{W/N},t\right).
\label{eq:trans01}
\end{equation}

Next we note that, while Eq.~(\ref{eq:fp2a}) is nonlinear, it is invariant under multiplicative scaling of the dependent variable.  It follows that the scaled dependent variable,
\begin{equation}
\overline{\overline{P}} = \frac{\overline{P}}{N}
\label{eq:trans2}
\end{equation}
will also solve Eq.~(\ref{eq:fp2a}), 
\begin{equation}
\frac{\partial \overline{\overline{P}}}{\partial t}
=
-\frac{\partial}{\partial \overline{w}}\left[\tau\left(1 - \overline{w}\right)\overline{\overline{P}}\right] +
\frac{\partial^2}{\partial {\overline{w}}^2}\left[\gamma\left(\overline{\overline{B}} + \frac{{\overline{w}}^2}{2}\overline{\overline{A}}\right) \overline{\overline{P}}\right],
\label{eq:fp2b}
\end{equation}
but will simplify the Pareto potentials so that they no longer involve $N$ at all,
\begin{eqnarray}
\overline{\overline{A}}(\overline{w},t)
& := &
\int_{\overline{w}}^{\infty} d\overline{x}\; \overline{\overline{P}}(\overline{x},t)\label{eq:AbarRef}\\
\overline{\overline{B}}(\overline{w},t)
& := &
\int_0^{\overline{w}} d\overline{x}\; \overline{\overline{P}}(\overline{x},t) \frac{{\overline{x}}^2}{2}.\label{eq:BbarRef}
\end{eqnarray}
In some sense, solving for $\overline{\overline{P}}$ is equivalent~\footnote{Since we introduced $N$ as the number of agents, it may seem confusing to set $N=1$.  The point of the continuum limit, however, is that the number of agents may be real-valued, and more naturally defined as the zeroth moment of the solution to the Fokker-Planck equation.  Hence, in the continuum limit, there is no reason not to take $N=1$ for our canonical form.} to solving the problem for $N=1$.  Since we have already scaled the independent variable so that $W/N=1$, it follows that $\overline{\overline{P}}$ is the solution for the agent density function when $N=W=1$.  Once that solution for $\overline{\overline{P}}$ is known, Eqs.~(\ref{eq:trans01}) and (\ref{eq:trans2}) tell us that
\begin{equation}
P(w,t) = \frac{N^2}{W}\overline{\overline{P}}\left(\frac{w}{W/N},t\right).
\end{equation}

Finally, if the transaction rate $\gamma$ is independent of $w$ and $t$, we can make one more transformation.  We can measure time on the transactional scale by defining the new independent variable
\begin{equation}
\overline{t} = \gamma t,
\label{eq:tTrans}
\end{equation}
and then we can define
\begin{equation}
\taubar = \tau/\gamma.
\end{equation}
Since $\tau$ measures redistribution per unit time, and $\gamma$ measures transactions per unit time, $\taubar$ is the level of redistribution per transaction.  With this transformation, the Fokker-Planck equation becomes
\begin{equation}
\frac{\partial \overline{\overline{P}}}{\partial \overline{t}}
=
-\frac{\partial}{\partial \overline{w}}\left[\taubar\left(1 - \overline{w}\right)\overline{\overline{P}}\right] +
\frac{\partial^2}{\partial {\overline{w}}^2}\left[\left(\overline{\overline{B}} + \frac{{\overline{w}}^2}{2}\overline{\overline{A}}\right) \overline{\overline{P}}\right].
\label{eq:fp2c}
\end{equation}
Again, this is equivalent to solving Eq.~(\ref{eq:fp2b}) for $\gamma=1$, and employing the transformation
\begin{equation}
P(w,t) = \frac{N^2}{W}\overline{\overline{P}}\left(\frac{w}{W/N},\gamma t\right)
\label{eq:transFinal}
\end{equation}
to recover the solution of the original equation.

Thus, the net effect of all these scaling transformations is to demonstrate that if we can solve the original Fokker-Planck equation, Eq.~(\ref{eq:fp1}) for $N=W=\gamma=1$, then we can solve it for any values of $N$, $W$ and $\gamma$.  Henceforth, we remove all of the bars over the variables except $\taubar$, which is the redistribution per transaction, and we focus on the case $N=W=\gamma=1$, which we refer to as the {\it canonical form} of Eq.~(\ref{eq:fp1}).  All of our analysis henceforth will be aimed at solving the canonical form of the equation with the single free parameter $\taubar$.

\subsection{Steady-state agent density function}
\label{ssec:sss}

To find the steady state of Eq.~(\ref{eq:fp1}) with $N=W=\gamma=1$, we set $\partial P/\partial t = 0$ and integrate once with respect to $w$ to obtain the first-order, nonlinear integrodifferential equation
\begin{equation}
\frac{d}{dw}\left[\left(B+\frac{w^2}{2}A\right)P\right]
=
\taubar\left(1 - w\right)P.
\label{eq:fp1ss}
\end{equation}
Here we have changed the derivative with respect to $w$ to a total derivative, since time $t$ is no longer an independent variable, and it is worth remembering that the functions $A$ and $B$ are given by
\begin{eqnarray}
A(w)
& := &
\int_w^{\infty} dx\; P(x)\label{eq:A1ss}\\
B(w)
& := &
\int_0^w dx\; P(x) \frac{x^2}{2},\label{eq:B1ss}
\end{eqnarray}
which are simply the steady-state versions of Eqs.~(\ref{eq:AbarRef}) and (\ref{eq:BbarRef}).  The numerical method used for finding the steady-state solutions of this canonical-form equation is presented in Appendix~\ref{sec:numerical}.

\begin{figure}
\begin{center}
\includegraphics[bbllx=0,bblly=0,bburx=339,bbury=270,width=.40\textwidth]{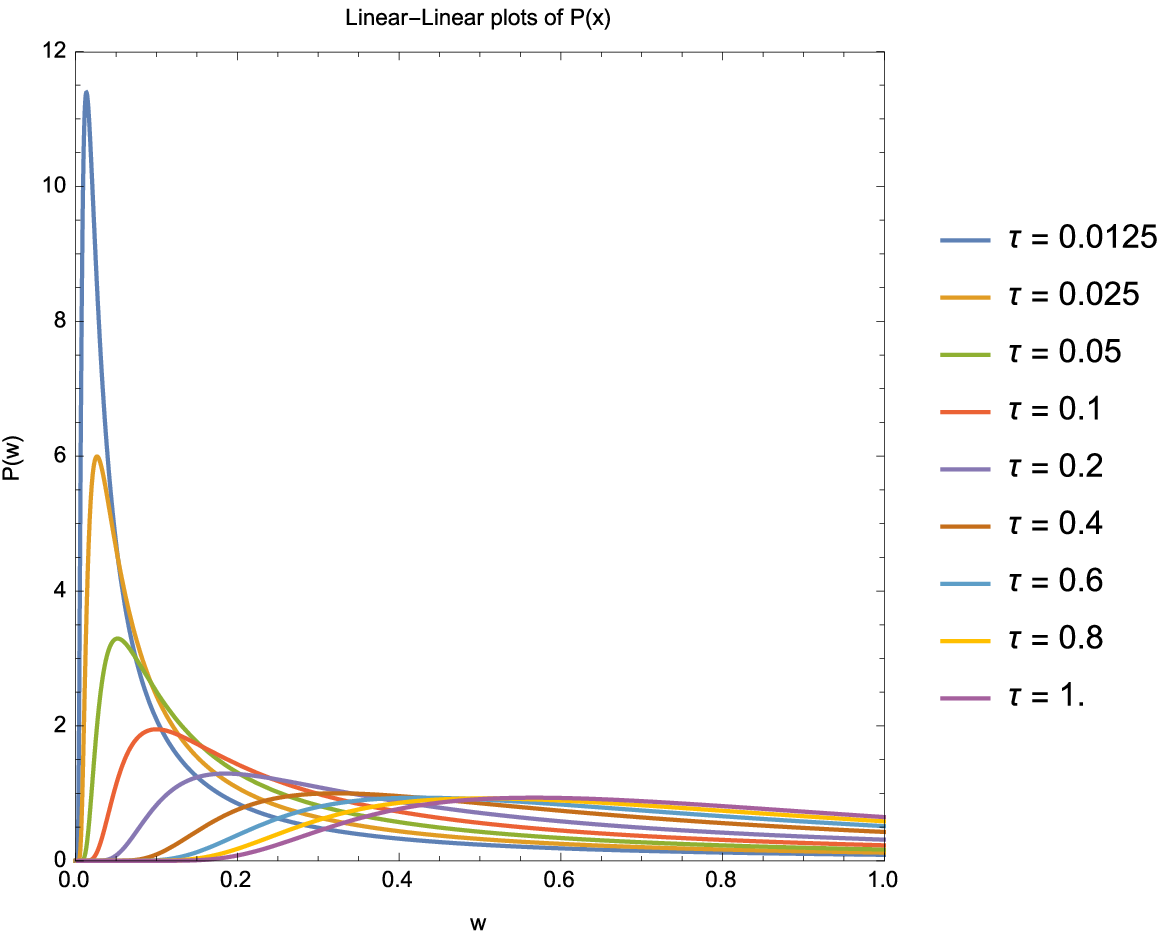}
\hspace{0.1in}
\includegraphics[bbllx=0,bblly=0,bburx=324,bbury=267,width=.40\textwidth]{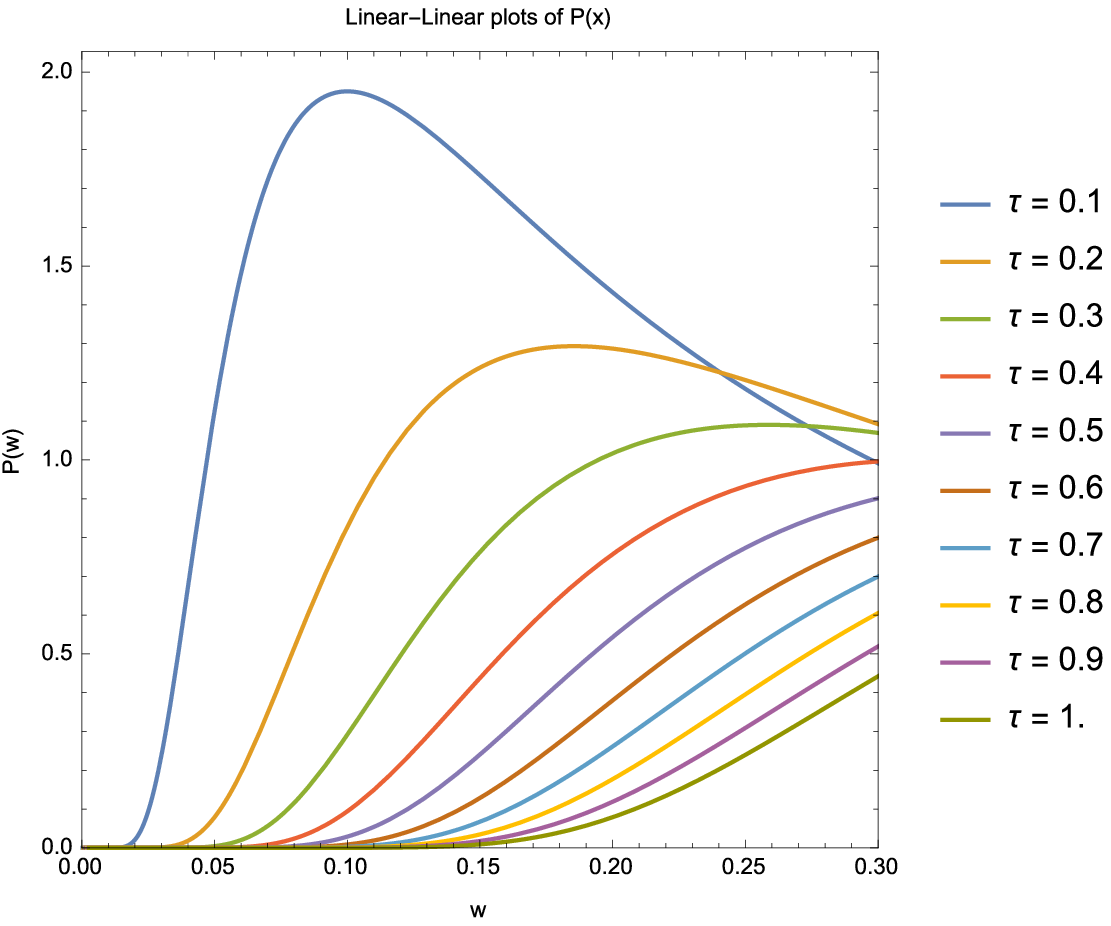}\\
\vspace{0.1in}
\includegraphics[bbllx=0,bblly=0,bburx=339,bbury=273,width=.40\textwidth]{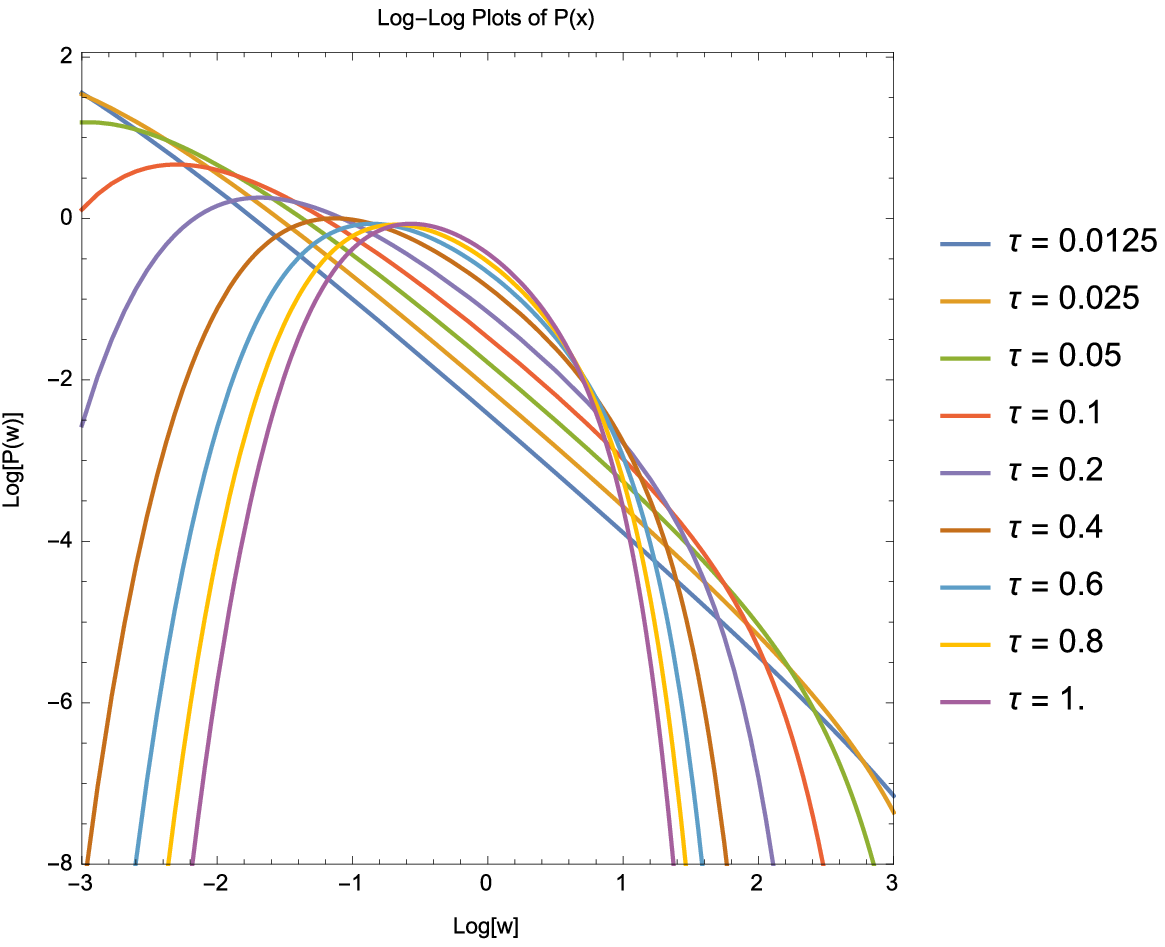}
\hspace{0.1in}
\includegraphics[bbllx=0,bblly=0,bburx=339,bbury=266,width=.40\textwidth]{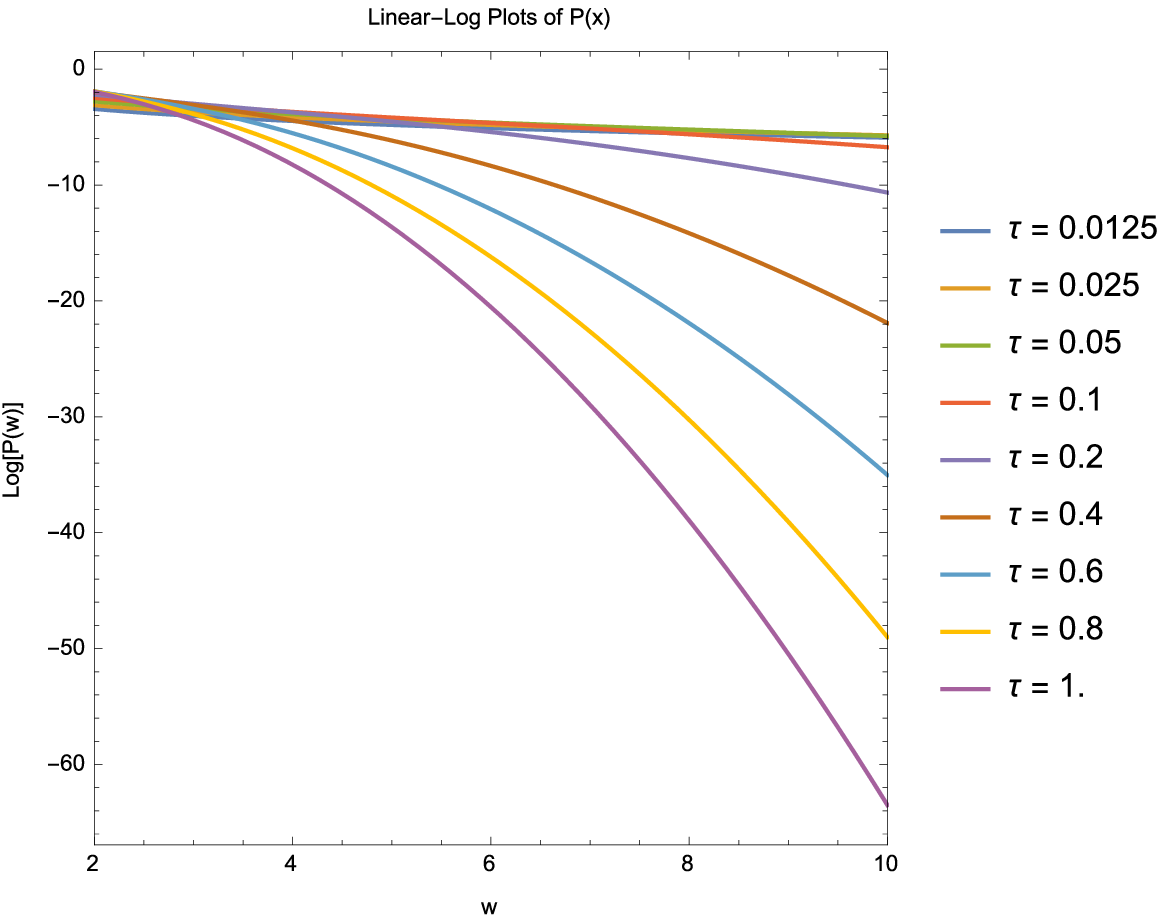}
\end{center}
\caption{{\bf Numerical solutions to Eq.~(\ref{eq:fp1ss}) for $W/N=1$ and various values of $\taubar$:} The upper left plot shows the steady-state distributions $\lim_{t\rightarrow\infty}P(w,t)$.  The region near the origin is magnified in the upper right plot, showing the depletion of these distributions near the origin, due to the redistribution.  The log-log plot on the lower left shows that the approximate power law when redistribution is low gives way to a lognormal distribution when the redistribution is increased.  Finally, the approximate parabolic curves in the linear-log plot on the lower right suggest that the tails at very large values of wealth are gaussian in nature.}
\label{fig:Dist.All}
\end{figure}

An illustration of numerical solutions to Eqs.~(\ref{eq:fp1ss}), (\ref{eq:A1ss}) and (\ref{eq:B1ss}) for various values of $\taubar$ is shown in the upper left plot of Fig.~\ref{fig:Dist.All}.  The other three plots in that figure emphasize three key features of these solutions.
\begin{itemize}
\item First, the curves are very flat in the vicinity of the origin.  For very low values of $\taubar$ this is barely visible in the upper left plot, but it becomes more noticeable at higher values of $\taubar$; the magnification in the upper right plot makes clear the presence of this flat region for all $\taubar$.
\item Second, after increasing to their maxima, the plots decay very differently depending on the magnitude of $\taubar$.  For very low values of $\taubar$, the plots decay approximately linearly on the log-log plot on the lower left of the figure, corresponding to a power law.  For the lowest values of $\taubar$ in the plot, this approximate linearity is sustained across six $e$ foldings of $w$; even for $\taubar$ as large as $0.1$, the approximate linearity is sustained over two $e$ foldings of $w$, from about $e^{-1}$ to $e^{+1}$.  For larger values of $\taubar$, and certainly for $\taubar \geq 0.4$, the plot is better approximated by an inverted parabola or catenary fitted to its maximum, corresponding to a lognormal or stretched-exponential distribution, respectively.
\item Third, for very large $w$ -- say, for $w > 4$ -- the plots are well approximated by downward-opening parabolas on the linear-log plot on the lower right of the figure, suggesting a gaussian tail.
\end{itemize}
The first two of these features are the defining characteristics of the famous Pareto law of wealth distribution, first posited by Pareto about a century ago.  These features were noted in earlier work on the YSM by Boghosian~\cite{bib:Boghosian1}, based on a number of numerical solutions of Eqs.~(\ref{eq:fp1ss}), (\ref{eq:A1ss}) and (\ref{eq:B1ss}) for various values of $\taubar$.  Since then, we have undertaken a much more extensive parametric study to obtain a fuller picture of the behavior of the solution to these equations in each of these three regimes, and we describe these results in the following subsections.

\subsubsection{Asymptotic analysis of steady state at low wealth}

To understand the very flat regions visible in the upper right plot of Fig.~\ref{fig:Dist.All} at low $w$, following an argument presented in earlier work~\cite{bib:Boghosian1}, we make the approximations $A\approx 1$ and $B\approx 0$ in Eq.~(\ref{eq:fp1ss}), which we shall justify a posteriori.  With these approximations, Eq.~(\ref{eq:fp1ss}) becomes
\begin{equation}
\frac{d}{dw}\left(\frac{w^2}{2}P\right)
=
\taubar\left(1 - w\right)P,
\label{eq:fp1ssLowW}
\end{equation}
which is linear and easily solved to yield
\begin{equation}
P(w)\approx\frac{C_0}{w^{2+2\taubar}}\exp\left(-\frac{2\taubar}{w}\right),
\label{eq:smallW}
\end{equation}
where $C_0$ is a constant of integration.  This function is non-analytic at the origin because all of its derivatives vanish there, hence its Taylor series vanishes at the origin, explaining the flatness observed near the origin in the graphs of the upper right-hand plot of Fig.~\ref{fig:Dist.All}.  Eqs.~(\ref{eq:A1ss}) and (\ref{eq:B1ss}) then imply that all the derivatives of $A$ and $B$ vanish at the origin.  Since $A(0)=1$ and $B(0)=0$, we have the a posteriori justification for the approximations that led to Eq.~(\ref{eq:fp1ssLowW}).

From the argument of the exponential factor in Eq.~(\ref{eq:smallW}), it is clear that the characteristic width of the flat region will increase with $\taubar$, consistent with what is observed in the upper right plot of Fig.~\ref{fig:Dist.All}.  As noted earlier, Pareto~\cite{bib:Pareto}, based on his observations of land ownership in Italy, Switzerland and Germany, actually conjectured a sharp cutoff -- i.e., that $P=0$ for $w < w_{\mbox{\tiny$\min$}}$.  We now see that the assumption of even a small level of redistribution in those countries at that time is sufficient to explain his observations within the context of the YSM.

\subsubsection{Numerical analysis of steady state at intermediate wealth}

The intermediate regime, in which $w$ is decreasing after its maximum, is perhaps the most difficult to analyze, because of the lack of any small parameter.  We describe our methodology for numerical solution of Eq.~(\ref{eq:fp1ss}) for the steady-state agent density function in Appendix~\ref{sec:numerical}.  The numerical solutions in the lower left plot in Fig.~\ref{fig:Dist.All} show that, for the very lowest values of $\taubar$, the log-log plots of $P$ are straight over six $e$ foldings.  [Note added in press:  In very recent work, Bustos Guajardo and Moukarzel~\cite{bib:BGM} have shown this limiting slope to be $-3/2$, consistent with Fig.~\ref{fig:Dist.All}.]  As noted earlier, this power-law decrease is a defining characteristic of Pareto distributions, and hence consistent with at least some empirical observations.  Pareto posited~\cite{bib:Pareto} that $A$ decays as $w^{-\alpha}$, and hence $P$ decays as $w^{-\alpha-1}$, where $\alpha$ is known as the {\it Pareto-Lorenz coefficient} and is widely measured and published.

The log-log plots in Fig.~\ref{fig:Dist.All} also show how this power-law dependence breaks down at higher values of $\taubar$.  They show clearly that, for approximately $\taubar > 0.4$, the functional form of $P$ becomes everywhere concave down on the log-log plot.  In fact, this effect was noted by Gibrat in 1931~\cite{bib:Gibrat}, who proposed a law which predicted a lognormal distribution for $P$,
\begin{equation}
P(w) \propto \frac{1}{w}\exp\left[-\beta^2 \log^2 \left(\frac{w}{w_0}\right)\right],
\end{equation}
where $\beta$ and $w_0$ are positive constants, and $\beta$ is called the {\it Gibrat index}~\footnote{This usage of $\beta$ is not to be confused with that in Section~\ref{sec:intro}}.

Our numerical study of the intermediate regime of the redistributive YSM suggests that Pareto's observations are more likely to be valid for societies with lower redistribution rates, while Gibrat's observations are more relevant for societies with higher redistribution rates.

\subsubsection{Asymptotic analysis of steady state at large wealth}

At very large values of $w$, a different analysis is warranted.  Because it is rather detailed, we relegate its  presentation to Appendix~\ref{sec:LargeW}.  The asymptotic analysis given there is valid for the more general model that includes Wealth-Attained Advantage (WAA), as will be discussed in Sec.~\ref{sec:waa}, but in the absence of WAA, the result reduces to the gaussian,
\begin{equation}
P(w) \approx C_\infty\exp\left[-\frac{\taubar}{2B_\infty} \left(w^2-2w\right)\right],
\label{eq:gaussian}
\end{equation}
where $B_\infty$ and $C_\infty$ are positive constants.

Eq.~(\ref{eq:gaussian}) is consistent with three features of the linear-log plot of $P(w)$ for large $w$ in the lower right of Fig.~\ref{fig:Dist.All}.  First, the graphs on that linear-log plot are approximately parabolic in nature.  Second, as was the case in the log-log plots of $P(w)$ for intermediate wealth, the rate of decay increases with the redistribution per transaction, $\taubar$.  Third, the gaussian fits to $P(w)$ for large $w$ all achieve their maximum at approximately the same location, namely $w=1$.

\subsection{Steady-state Gini coefficient}

\begin{figure}
\begin{center}
\includegraphics[bbllx=0,bblly=0,bburx=398,bbury=273,width=.50\textwidth]{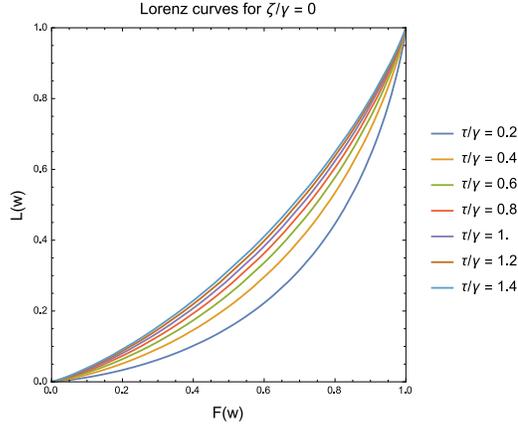}
\end{center}
\caption{{\bf Lorenz curves} for various values of $\taubar$, demonstrating that the Gini coefficient decreases with increasing $\taubar$.}
\label{fig:LorenzPlots0}
\end{figure}

\begin{figure}
\begin{center}
\includegraphics[bbllx=0,bblly=0,bburx=332,bbury=190,width=.50\textwidth]{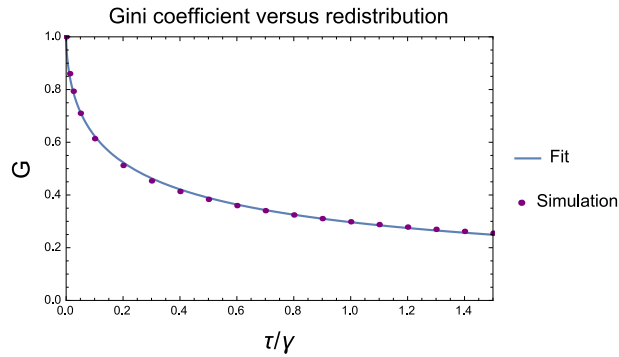}
\end{center}
\caption{{\bf Time asymptotic Gini coefficient ${\mathbf G_\infty}$} versus $\taubar$ for $W/N=1$.  The data points were obtained from numerical solution of Eq.~(\ref{eq:fp2}), while the solid curve is the fit given in Eq.~(\ref{eq:GFit}).}
\label{fig:GiniVsTaubar}
\end{figure}

Lorenz curves for steady-state solutions of the redistributive YSM are plotted in Fig.~\ref{fig:LorenzPlots0}, illustrating that the time-asymptotic Gini coefficient
\begin{equation}
G_\infty := \lim_{t\rightarrow\infty} G(t)
\end{equation}
decreases with $\taubar$.  The relationship between $G_\infty$ and $\taubar$ for $W/N=1$, is then plotted in Fig.~\ref{fig:GiniVsTaubar}.  The data points were obtained by numerical solution of Eq.~(\ref{eq:fp2}), while the solid curve is the remarkably accurate fit to these points,
\begin{equation}
G_\infty \approx \frac{1}{1 + 2.3682\;{\taubar}^{0.5943}}.
\label{eq:GFit}
\end{equation}

From Fig.~\ref{fig:GiniVsTaubar} it is evident that $G_\infty$ decreases with the redistribution rate rather sharply for $G_\infty$ greater than about $0.5$, but only painfully slowly for $G_\infty$ less than about $0.25$.  For example, according to Eq.~(\ref{eq:GFit}), it takes about a factor of $6.351$ increase in the redistribution per transaction, $\taubar$, to lower $G_\infty$ from $0.50$ to $0.25$.  Yet another factor of $6.351$ increase would be required to lower $G_\infty$ further from $0.25$ to $0.10$.  We delay discussion of the implications of this observation until after the introduction of wealth-attained advantage in the following section.

Finally, we note that the Gini coefficient is not a Lyapunov functional of Eq.~(\ref{eq:fp2}).  It can not be, because it is possible to begin with initial condition $G(0) > G_\infty$, in which case the Gini coefficient would have to decrease in time.  To date, no Lyapunov functional is known for Eq.~(\ref{eq:fp2}) that achieves a maximum for a solution of Eq.~(\ref{eq:fp1ss}).

\section{Wealth-attained advantage in the Yard-Sale Model}
\label{sec:waa}
\subsection{Derivation of the Fokker-Planck equation for the redistributional YSM with WAA}

To introduce Wealth-Attained Advantage (WAA) in the YSM, we need to bias the outcome of the coin flip that determines the direction of wealth transfer in a transaction.  We continue to suppose that an agent with wealth $\overline{w}$ engages in a transaction with an agent with wealth $\overline{x}$.  We retain Eq.~(\ref{eq:deltaWRedist}) for $\Delta w$.  From Eq.~(\ref{eq:transform}), we see that $\eta=+1$ corresponds to this amount flowing to the agent with starting wealth $\overline{w}$, and $\eta=-1$ corresponds to this amount flowing from that agent.  Since $\eta\in\{-1,+1\}$, we still have $E[\eta^2]=1$, but we change Eq.~(\ref{eq:Eeta}) for the expectation value of $\eta$ to read
\begin{equation}
E[\eta] = \zeta\;\frac{N}{W}\sqrt{\frac{\Delta t}{\gamma}}\;\left(\overline{w}-\overline{x}\right).
\label{eq:EetaWAA}
\end{equation}
This means that the probability that $\Delta w$ flows to the agent with wealth $\overline{w}$ is proportional to the difference $\overline{w}-\overline{x}$.  The proportionality constant is controlled by the new variable $\zeta$; the other factors in this proportionality constant are intended only to simplify the resulting Fokker-Planck equation.

With the new definition of the expectation value as defined above, the new drift term in the Fokker-Planck equation is then
\begin{equation}
\sigma = \calE\left[\frac{\Delta w}{\Delta t}\right] = \tau\left(\frac{W}{N} - w\right)
-2\zeta\left[\frac{N}{W}\left(B - \frac{w^2}{2}A\right) + w\left(\frac{1}{2} - L\right)\right],
\end{equation}
while the diffusivity given in Eq.~(\ref{eq:D1}) is unchanged.  The nonlinear, integrodifferential Fokker-Planck equation for the redistributive YSM with WAA is therefore
\begin{equation}
\boxed{
\frac{\partial P}{\partial t}
=
-\frac{\partial}{\partial w}\left[\tau\left(\frac{W}{N}-w\right)P\right]
+\frac{\partial}{\partial w}
\left\{\zeta\left[2\frac{N}{W}\left(B-\frac{w^2}{2}A\right) + \left(1-2L\right)w\right]P\right\} +
\frac{\partial^2}{\partial w^2}\left[\gamma\left(B + \frac{w^2}{2}A\right)P\right].}
\label{eq:fp3}
\end{equation}
The second term on the right models the presence of WAA, as measured by $\zeta$.

That Eq.~(\ref{eq:fp3}) still conserves agents is manifest.  That it still conserves wealth follows from an integration by parts, from Eq.~(\ref{eq:ConsId1}), and from the identity
\begin{equation}
0 = \int_0^\infty dw\; \left[2\frac{N}{W}\left(B-\frac{w^2}{2}A\right) + \left(1-2L\right)w\right]P.
\label{eq:ConsId2}
\end{equation}
Eq.~(\ref{eq:ConsId2}) may be verified by using the definitions of $A$, $B$ and $L$, given in Eqs.~(\ref{eq:A}), (\ref{eq:B}) and (\ref{eq:L}) respectively, and then judiciously reversing the order of integration.

\subsection{Scaling and canonical form}

As in Subsection~\ref{ssec:scaling}, we can transform the independent variable using Eqs.~(\ref{eq:trans0}) and (\ref{eq:tTrans}), and the dependent variable using Eq.~(\ref{eq:trans2}).  Though it requires some additional work to do this for the WAA term, just as was shown in that subsection, the final result is equivalent to solving a canonical form of Eq.~(\ref{eq:fp3}) with $N=W=\gamma=1$ for $\overline{\overline{P}}(\overline{w},\overline{t})$, and then recovering the solution for the desired values of $N$, $W$ and $\gamma$ from Eq.~(\ref{eq:transFinal}).  This time, we additionally define the WAA per transaction,
\begin{equation}
\zetabar = \zeta/\gamma,
\end{equation}
and henceforth we drop the bars over all the variables except $\taubar$ and $\zetabar$.

The resulting equation in canonical form is
\begin{equation}
\frac{\partial P}{\partial t}
=
-\frac{\partial}{\partial w}\left[\taubar\left(1-w\right)P\right]
+\frac{\partial}{\partial w}
\left\{\zetabar\left[2\left(B-\frac{w^2}{2}A\right) + \left(1-2L\right)w\right]P\right\} +
\frac{\partial^2}{\partial w^2}\left[\left(B + \frac{w^2}{2}A\right)P\right],
\label{eq:fp3Canonical}
\end{equation}
where the Pareto potentials are given by
\begin{eqnarray}
A(w,t)
& := &
\int_{w}^{\infty} dx\; P(x,t)
\label{eq:ACanonical}\\
L(w,t)
& := &
\int_{0}^{w} dx\; P(x,t) x
\label{eq:LCanonical}\\
B(w,t)
& := &
\int_0^{w} dx\; P(x,t) \frac{x^2}{2}.
\label{eq:BCanonical}
\end{eqnarray}

\subsection{Steady-state agent density function}
\label{ssec:sssWAA}

As in Subsection~\ref{ssec:sssWAA}, we can find an equation for the steady-state solutions of Eqs.~(\ref{eq:fp3Canonical}) through (\ref{eq:BCanonical}) by setting the time derivative of $P$ to zero, and integrating once with respect to $w$ to obtain the ordinary differential equation
\begin{equation}
\frac{d}{dw}\left[\left(B + \frac{w^2}{2}A\right)P\right]
=
\taubar\left(1-w\right)P
-\zetabar\left[2\left(B-\frac{w^2}{2}A\right) + \left(1-2L\right)w\right]P.
\label{eq:fp3CanonicalSS}
\end{equation}
where the Pareto potentials in steady state are functions only of $w$,
\begin{eqnarray}
A(w)
& := &
\int_{w}^{\infty} dx\; P(x)
\label{eq:ACanonicalSS}\\
L(w)
& := &
\int_{w}^{\infty} dx\; P(x) x
\label{eq:LCanonicalSS}\\
B(w)
& := &
\int_0^{w} dx\; P(x) \frac{x^2}{2}.
\label{eq:BCanonicalSS}
\end{eqnarray}
These equations admit a two-parameter family of solutions, where the parameters are $\taubar=\tau/\gamma$ which measures the level of redistribution per transaction, and $\zetabar=\zeta/\gamma$ which measures the level of WAA per transaction.

\subsubsection{Asymptotic analysis of steady state at low wealth}

To understand the behavior of solutions to Eq.~(\ref{eq:fp3CanonicalSS}) in the vicinity of the origin, we proceed in fashion similar to that in Subsection~\ref{ssec:sss}.  Make the approximations $A\approx 1$, $L\approx 0$ and $B\approx 0$, which will have to be justified a posteriori.  Then Eq.~(\ref{eq:fp3CanonicalSS}) reduces to
\begin{equation}
\frac{d}{dw}\left(\frac{w^2}{2}P\right)
=
\taubar\left(1-w\right)P
-\zetabar\left(w-w^2\right)P,
\end{equation}
which is linear and easily solved to yield
\begin{equation}
P(w)\approx\frac{C_0}{w^{2+2\taubar+2\zetabar}}\exp\left(-\frac{2\taubar}{w}+2\zetabar w\right).
\label{eq:fp1ssLowWWAA}
\end{equation}
It is clear that this reduces to Eq.~(\ref{eq:smallW}) when $\zetabar=0$, and hence is a generalization of that result.  It is also straightforward to see that it preserves the property of being non-analytic at the origin because all its derivatives vanish there.  It follows from Eqs.~(\ref{eq:ACanonicalSS}), (\ref{eq:LCanonicalSS}) and (\ref{eq:BCanonicalSS}) that all the derivatives of $A$, $L$ and $B$ also vanish at the origin.  Since $A(0)=1$ and $L(0)=B(0)=0$, we have the a posteriori justification for the approximations that led to Eq.~(\ref{eq:fp1ssLowWWAA}).

The above argument shows that the depletion of the distribution near the origin, consistent with Pareto's observations, survives the introduction of WAA.  Also, the argument of the exponential factor in Eq.~(\ref{eq:fp1ssLowWWAA}) makes evident that the width of this depleted region will increase with $\taubar$, just as it did in the absence of WAA.

\subsubsection{Numerical analysis of steady state at intermediate wealth:  Evidence of criticality}

Using the numerical method described in Appendix~\ref{sec:numerical}, we searched for solutions of Eqs.~(\ref{eq:fp3CanonicalSS}) through (\ref{eq:BCanonicalSS}).  The shooting method used begins with the initial conditions $A(0)=1$ and $P(0)=L(0)=B(0)=0$, and searches for a solution with $\lim_{w\rightarrow\infty}A(w)=0$.  For $\zetabar < \taubar$ and intermediate $w$, the solutions found were similar in form to those depicted in Fig.~\ref{fig:Dist.All}, with an approximate power law at low $\taubar$ and an approximate gaussian at high $\taubar$.

\begin{figure}
\begin{center}
\includegraphics[bbllx=0,bblly=0,bburx=454,bbury=185,width=.65\textwidth]{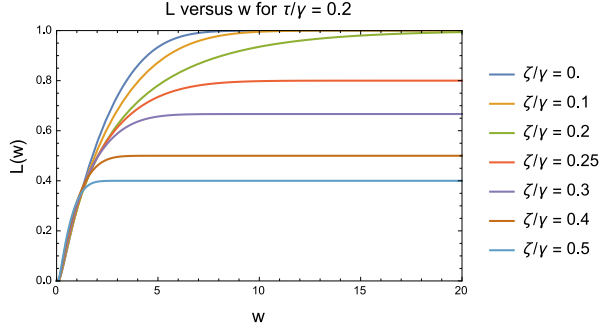}
\end{center}
\caption{{\bf The Pareto potential $\mbox{\boldmath $L(w)$}$} versus $w$ for $\taubar=0.2$ and various values of $\zetabar$.  The change in behavior for $\zetabar$ above and below criticality is evident.  Below criticality, the limiting value of $L$ is one and the solution is not oligarchical; above criticality, the limiting value of $L$ is less than one, and the remaining fraction of wealth is tied up in the oligarchical term.}
\label{fig:LVsW}
\end{figure}

For $\zetabar > \taubar$, we found that the solutions abruptly changed character.  The best way to appreciate this change is to study $\lim_{w\rightarrow\infty}L(w)$ as a function of $w$, as shown for $\taubar=0.2$ and various values of $\zetabar$ in Fig.~\ref{fig:LVsW}.  More generally, to within eight significant digits, our numerical solutions indicate
\begin{equation}
\lim_{w\rightarrow\infty} L(w) =
\left\{
\begin{array}{ll}
1 & \mbox{for $\zetabar < \taubar$}\\
\frac{\taubar}{\zetabar} & \mbox{for $\zetabar > \taubar$.}
\end{array}
\right.
\label{eq:LFinalGen}
\end{equation}

The fact that $L(w)$ does not approach one as $w$ approaches infinity is the signature of an oligarchical solution, of the form of Eq.~(\ref{eq:oligarchicalSol}) in Section~\ref{ssec:Xi}, with
\begin{equation}
W_\Xi = 
\left\{
\begin{array}{ll}
0 & \mbox{for $\zetabar < \taubar$}\\
1-\frac{\taubar}{\zetabar} & \mbox{for $\zetabar > \taubar$.}
\end{array}
\right.
\end{equation}
In other words, a fraction $1-\taubar/\zetabar$ of the total wealth is ``at infinity,'' or ``held by the oligarch,'' and therefore does not appear in the limit $\lim_{w\rightarrow\infty}L(w)$.

\begin{figure}
\begin{center}
\includegraphics[bbllx=0,bblly=0,bburx=390,bbury=271,width=.50\textwidth]{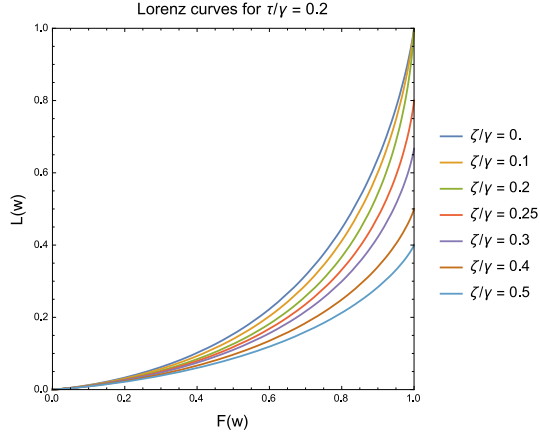}
\end{center}
\caption{{\bf Lorenz curves} for $\taubar=0.2$ and various values of $\zetabar$.  The change in behavior for $\zetabar$ above and below criticality is evident.  Below criticality, the Lorenz curves terminate at the point $(1,1)$; above criticality, they terminate at a value of $L$ less than unity.}
\label{fig:LorenzPlots}
\end{figure}

Another way to see the oligarchical character of the solution for $\zetabar > \taubar$ is to examine Lorenz curves; these are shown for $\taubar=0.2$ and various values of $\zetabar$ in Fig.~\ref{fig:LorenzPlots}.  We see that for $\zetabar < \taubar$ the termini of the Lorenz curve are $(0,0)$ and $(1,1)$.  For $\zetabar > \taubar$, however, the termini are $(0,0)$ and $(1,\taubar/\zetabar)$.  Again, this indicates that a fraction $1-\taubar/\zetabar$ of the total wealth has ``condensed'' into the hands of a vanishingly small fraction of the population.

\subsubsection{Asymptotic analysis of steady state at large wealth:  More evidence of criticality}

Another phenomenon associated with the critical point $\zetabar=\taubar$ can be appreciated by analyzing the tail of the agent density function $P(w)$, for very large values of $w$.  This analysis is presented in Appendix~\ref{sec:LargeW}, and it yields a striking result.  We find that the character of the tail changes abruptly as $\zetabar$ is increased past $\taubar$.  The asymptotic form for large $w$ is
\begin{equation}
P(w)\approx
\left\{
\begin{array}{lll}
C_\infty\exp\left[-\left(\frac{\taubar-\zetabar}{2B_\infty}\right) w^2 - \left(\frac{2B_\infty\zetabar-\taubar}{B_\infty}\right) w\right] & & \mbox{for $\zetabar \leq \taubar$}\\ \\
C_\infty\exp\left[-\left(\frac{\zetabar-\taubar}{2B_\infty}\right) w^2 - \left(\frac{2B_\infty\zetabar-\taubar}{B_\infty}\right) w\right] + \left(1 - \frac{\taubar}{\zetabar}\right)\Xi(w) & & \mbox{for $\zetabar > \taubar$.}
\end{array}
\right.
\label{eq:gaussianWAA}
\end{equation}
For the subcritical case $\zetabar=0$, the above reduces to Eq.~(\ref{eq:gaussian}), as expected.  As $\zetabar$ is increased from zero, the tail of the distribution remains asymptotically gaussian, until the critical point $\zetabar=\taubar$, where it degenerates to an exponential.  If we continue to increase $\zetabar$ above this critical value, the gaussian character of the tail returns, and the oligarchical term appears, as noted in the numerical solutions for intermediate $w$.

\subsection{Steady-state Gini coefficient}

We have solved Eq.~(\ref{eq:fp3CanonicalSS}) for various values of $\taubar$ and $\zetabar$, using the shooting method described in Appendix~\ref{sec:numerical}.  Fig.~\ref{fig:GiniVsTauAndZeta} shows the time-asymptotic value of the Gini coefficient, $G_\infty$, as a function of $\taubar$ and $\zetabar$, as a three-dimensional plot.  The red curve represents criticality, and the region to its upper left is the supercritical region where the solutions are oligarchical in nature.  The cross section with the plane $\zetabar=0$ at the very front of the figure is identical to the curve presented in Fig.~\ref{fig:GiniVsTaubar}.
\begin{figure}
\begin{center}
\includegraphics[bbllx=0,bblly=0,bburx=360,bbury=345,width=.40\textwidth]{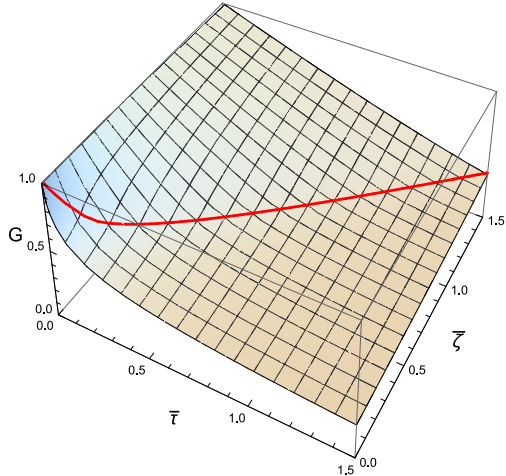}
\end{center}
\caption{{\bf Time asymptotic Gini coefficient ${\mathbf G_\infty}$} as a function of $\taubar$ and $\zetabar$, presented as a three-dimensional plot.  The red curve represents criticality, where the surface has a slope discontinuity (more visible in Fig.~\ref{fig:GiniVsTauManyZeta}), and the region to its upper left is the supercritical region where the solutions are oligarchical in nature.}
\label{fig:GiniVsTauAndZeta}
\end{figure}

\begin{figure}
\begin{center}
\includegraphics[bbllx=0,bblly=0,bburx=385,bbury=198,width=.60\textwidth]{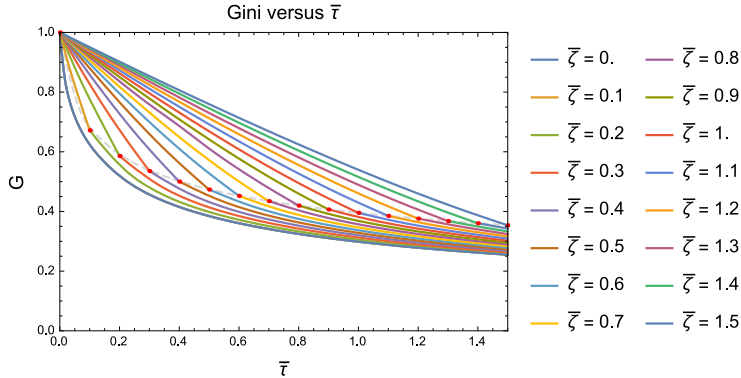}
\end{center}
\caption{{\bf Time asymptotic Gini coefficient ${\mathbf G_\infty}$} as a function of $\taubar$ for many different values of $\zetabar$, presented as superposed two-dimensional plots.  The red dots and gray dashed curve represent criticality.  The slope discontinuity at criticality is evident.}
\label{fig:GiniVsTauManyZeta}
\end{figure}

The surface depicted in Fig.~\ref{fig:GiniVsTauAndZeta} has a slope discontinuity at criticality -- that is, along the red curve in that figure -- but it is not so very evident in that figure.  Fig.~\ref{fig:GiniVsTauManyZeta}, which plots $G_\infty$ versus $\taubar$ for many different values of $\zetabar$, shows this discontinuity much more clearly.

From Fig.~\ref{fig:GiniVsTauManyZeta}, it is evident that WAA can have a substantial impact on the Gini coefficient if $\taubar$ is small, but has very little impact if $\taubar$ is large.  This suggests that societies that are more inclined to higher redistribution and lower Gini coefficients are also more robust to the effects of WAA.  Conversely, societies with lower redistribution and higher Gini coefficients are more sensitive to the effects of WAA.  This suggests a more quantitative way to frame a possible correlation between lower redistribution and a more stratified and entrenched economic class structure, but we leave this matter for public policy specialists to work out.

Of the nearly 200 countries in the world today, the vast majority have wealth Gini coefficients between 0.60 and 0.80.  A few outliers such as Japan and China have Gini coefficients slightly lower than this range, and others such as the United States and Zimbabwe have Gini coefficients slightly higher than this range~\cite{bib:GiniByCountry}.  By contrast, recent studies of the few remaining extant hunter-gatherer societies in the world today indicate that most have wealth Gini coefficients of 0.25 or below, and at least two of them actually have wealth Gini coefficients under 0.10~\cite{bib:Smith2010,bib:BorgerhoffMulder2009}.  In light of this data, the phase transition illustrated in Figs.~\ref{fig:GiniVsTauAndZeta} and \ref{fig:GiniVsTauManyZeta} takes on particular significance.  While it is difficult to relate $\overline{\tau}$ to actual data, the figures make clear that unless redistribution is very low indeed, a wealth Gini coefficient in excess of 0.6 probably indicates supercriticality and hence the presence of wealth condensation.  Conversely, unless redistribution is very high, a wealth Gini coefficient below 0.3 probably indicates subcriticality and hence the absence of wealth condensation.  It may be the case, therefore, that the phase transition identified in this paper is precisely what separates the economic phenomenology of industrialized and hunter-gatherer societies.

\begin{figure}
\begin{center}
\includegraphics[bbllx=0,bblly=0,bburx=618,bbury=516,width=.60\textwidth]{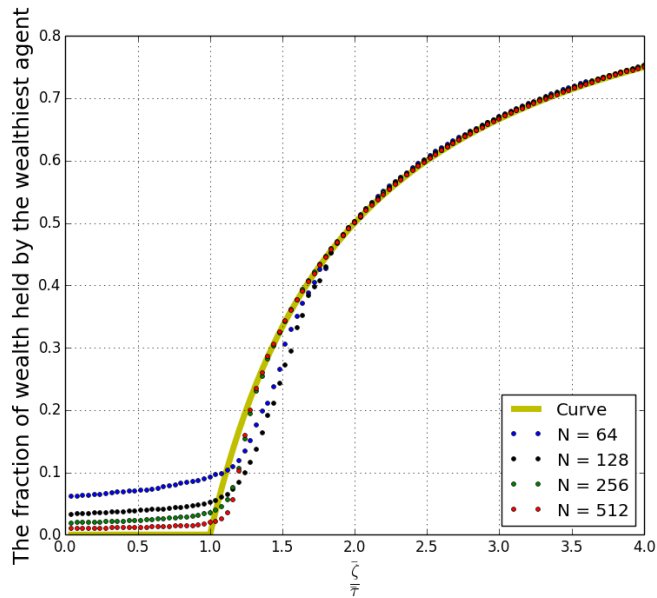}
\end{center}
\caption{{\bf Results of Monte Carlo simulations} for the fraction of wealth held by the wealthiest agent as a function of $\zetabar/\taubar$ for 64, 128, 256 and 512 agents.  The solid yellow-green curve is the theoretical result given by Eq.~(\ref{eq:fowhbwa}).  The critical point is clearly visible when $\zetabar/\taubar=1$, and the continuum limit with a slope discontinuity at $\zetabar/\taubar=1$ is approached as the number of agents increases.}
\label{fig:MC}
\end{figure}

Another method of visualizing the abrupt change that takes place at criticality is to employ a Monte Carlo simulation with a finite number of agents.  In that setting, the wealthiest agent plays the role of the oligarch, and the fraction of wealth held by that agent is directly observable, and ought to be given by
\begin{equation}
W_{\Xi,\infty} := \lim_{t\rightarrow\infty}W_\Xi(t) =
\left\{
\begin{array}{ll}
0 & \mbox{for $\zetabar\leq\taubar$}\\
1-\frac{\taubar}{\zetabar} & \mbox{for $\zetabar\geq\taubar$},
\end{array}
\right.
\label{eq:fowhbwa}
\end{equation}
which is shown as the solid yellow-green curve in Fig.~\ref{fig:MC}.  Superposed on this curve in the figure are the results of Monte Carlo simulations with 64, 128, 256 and 512 agents.  It is seen that there are finite-size effects for small numbers of agents, but these diminish as the number of agents is increased, and the curve approaches a continuum limit with a slope discontinuity at $\zetabar/\taubar=1$, as expected.

Finally, it bears mentioning that, although all of the above results displayed for $G_\infty$, $W_{\Xi,\infty}$ and $P(w)$ are for the time-asymptotic limit, the underlying theory is very much a time-dependent theory.  It may well be that the time derivative terms are important for modern economies that one might wish to study.  The world's wealth distribution today is known to be very dynamic in nature.  Still, even when conducting time-dependent studies, it is useful to understand the time asymptotic behavior of the theory under consideration, and we hope that the above parametric study will be helpful to future investigations.

\section{Conclusions}

We have extended the Yard-Sale Model of Asset exchange by introducing a new model of Wealth-Attained Advantage (WAA), wherein the level of bias favoring the wealthier agent in a transaction is a smooth function of the difference in wealth between the two interacting agents.  Prior studies have taken this dependence to be discontinuous, and have observed a first-order phase transition to a fully wealth-condensed state.  In contrast, the smoothness of this dependence in our model results in a second-order phase transition to a state of coexistence between wealth-condensed and non-wealth-condensed phases.

In the course of our analysis, we derived a nonlinear, integrodifferential Fokker-Planck equation corresponding to the model, and demonstrated its universality in a particular sense.  We presented a parametric study of the steady-state solutions of this equation as a function of the wealth redistribution rate $\taubar$ and the level of WAA $\zetabar$.  The above-mentioned second-order phase transition occurs at the critical value $\zetabar=\taubar$, where order parameters, such as the fraction of wealth held by the wealthiest agent and the Gini coefficient, exhibit a slope discontinuity.

Finally, we also found that the onset of wealth condensation has a reciprocal effect on the asymptotic behavior of the tail of the non-oligarchical portion of the wealth distribution.  We presented both theoretical and numerical evidence that, whereas this behavior is gaussian both below and above criticality, it degenerates to exponential decay precisely at criticality.

We believe that the results from this investigation indicate the applicability of the Yard-Sale Model to real-world economies, though many additional improvements to the model will probably be necessary before it can be used for public policy studies.  In particular, it will be necessary to add terms for production and consumption, as well as for more sophisticated models of redistribution.  Still, these results encourage us to believe that future variants of the Yard-Sale Model may provide the long-sought key to a fundamental understanding of the distribution of wealth.

\section*{Acknowledgements}

Two of us (BMB and MJ) would like to acknowledge the hospitality of the Economic Research Group at the Central Bank of Armenia in Dilijan, Armenia during the late stages of this work in the summer of 2015.  We would like to thank Sam Bowles for pointing out the relevance of our results to the Gini coefficients of hunter-gatherer societies~\cite{bib:Smith2010,bib:BorgerhoffMulder2009}.  We would like to thank Cristian Moukarzel and Ricardo Bustos Guajardo for their helpful comments on our preprint after it was released on ArXiv; they had been working along similar lines~\cite{bib:BGM}, had independently anticipated our asymptotic analysis demonstrating the gaussian behavior of the tail, and had cleverly demonstrated that the slope of the lowest curve in the lower left plot of our Fig.~\ref{fig:Dist.All} is $-3/2$.  Helpful conversations are also acknowledged with Jie Li, Peter Love and Jan Tobochnik.

\bibliographystyle{plain.bst}
%\bibliography{paper.bib}

\begin{thebibliography}{}

\bibitem{bib:Boghosian1}{
B.M.~Boghosian,
{``Kinetics of wealth and the Pareto law''},
{\it Physical Review E}
{\bf 89}
(2014)
{pp. 042804--042825}.
}
\bibitem{bib:Boghosian2}{
B.M.~Boghosian,
{``Fokker-Planck description of wealth dynamics and the origin of Pareto's law''},
{\it International Journal of Modern Physics C}
{\bf 25}
(2014)
{pp. 1441008--1441015}.
}
\bibitem{bib:Boghosian3}{
B.M.~Boghosian, M. Johnson, J.A. Marcq,
{``An H theorem for Boltzmann's equation for the Yard-Sale Model of asset exchange''},
{\it J. Stat. Phys.},
{\bf 161}
(2015)
{pp. 1339--1350}.
}
\bibitem{bib:LevySolomon1997}{
M. Levy, S. Solomon,
{``New evidence for the power-law distribution of wealth''},
{\it Physica A}
{\bf 242}
(1997)
{pp. 90--94}.
}
\bibitem{bib:Moukarzel2007}{
C.F. Moukarzel, S. Gon\c{c}alves, J.R. Iglesias, R. Huerta-Quintanilla,
{``Wealth condensation in a multiplicative random asset exchange model''},
{\it Europhys. J. Spec. Topics}
{\bf 143}
(2007)
{pp. 75--79}.
}
\bibitem{bib:Angle}{
J.~Angle,
{``The surplus theory of social stratification and the size distribution of personal wealth''},
{\it Social Forces}
{\bf 65}
(1986)
{pp. 293--326}.
}
\bibitem{bib:IspolatovKrapivskyRedner}{
S.~Ispolatov, P.L.~Krapivsky, S.~Redner,
{``Wealth distributions in asset exchange models''},
{\it The European Physical Journal B -- Condensed Matter}
{\bf 2}
(1998)
{pp. 267--276}.
}
\bibitem{bib:Chakraborti2002}{
A.~Chakraborti,
{``Distributions of money in model markets of economy''},
{\it Int. J. Mod. Phys. C}
{\bf 13}
(2002)
{pp. 1315--1321}.
}
\bibitem{bib:Hayes}{
B.~Hayes,
{``Follow the money''},
{\it American Scientist}
{\bf 90}
(2002)
{pp. 400--405}.
}
\bibitem{bib:RMJ}{
M.N. Rosenbluth, W.M. MacDonald, D.L. Judd,
{``Fokker-Planck Equation for an Inverse-Square Force''}
{\it Phys. Rev.}
{\bf 107}
(1957)
pp. 1--6.
}
\bibitem{bib:OU}{
G.E. Uhlenbeck, L.S. Ornstein,
{``On the theory of Brownian Motion''},
{\it Phys. Rev.}
{\bf 36}
(1930)
pp. 823--841.
}
\bibitem{bib:Pareto}{
V.~Pareto,
{``La Courbe de la Repartition de la Richesse"} (1896),
in G.~Busino, editor,
{``Oevres Completes de Vilfredo Pareto''},
Geneva: Librairie Droz.
(1965) 
pp. 1–5.
}
\bibitem{bib:BouchaudMezard2000}{
J.-P. Bouchaud, M. M\'{e}zard,
{``Wealth condensation in a simple model of the economy''},
{\it Physica A}
{\bf 282}
(2000)
pp. 536--545.
}
\bibitem{bib:BurdaJohnstonEtAl2002}{
Z. Burda, D. Johnston, J. Jurkiewicz, M. Kami\'{n}ski, M.A. Nowak, G. Papp, I. Zahed,
{``Wealth condensation in Pareto macroeconomies''},
{\it Phys. Rev. E}
{\bf 65}
(2002)
p. 026102.
}
\bibitem{bib:ExtremePoverty}{
M.~Cruz, J.~Foster, B.~Quillin, P.~Schellekens,
{``Ending Extreme Poverty and Sharing Prosperity:  Progress and Policies''},
{World Bank Group Policy Research Note}
{PRN/15/03}
(October 2015).
}
\bibitem{bib:BGM}{
R.~Bustos~Guajardo, C.F.~Moukarzel,
{``Wealth distribution under Yard-Sale exchange with proportional taxes''},
preprint and private communication
(2015).
}
\bibitem{bib:Gibrat}{
R.~Gibrat,
{``Les In\'{e}galit\'{e}s \'{e}conomiques''},
{Paris, France},
(1931).
}
\bibitem{bib:GiniByCountry}{
{``List of countries by distribution of wealth''},
{Wikipedia},\\
\texttt{https://en.wikipedia.org/wiki/List\char`_of\char`_countries\char`_by\char`_distribution\char`_of\char`_wealth}
(2015).
}
\bibitem{bib:Smith2010}{
E.A. Smith, K. Hill, F. Marlowe, D. Nolin, P. Wiessner, M. Gurven, S. Bowles, M. Borgerhoff Mulder, T. Hertz,  A. Bell,
{``Wealth Transmission and Inequality Among Hunter-Gatherers''},
{\it Curr Anthropol.}
{\bf 51}
(February 2010)
pp. 19-34.
}
\bibitem{bib:BorgerhoffMulder2009}{
M. Borgerhoff Mulder, S. Bowles, T. Hertz, A. Bell, J. Beise, G. Clark, I. Fazzio, M. Gurven, K. Hill, P.L. Hooper, W. Irons, H. Kaplan, D. Leonetti, B. Low, F. Marlowe, R. McElreath, S. Naidu, D. Nolin, P. Piraino, R. Quinlan, E. Schniter, R. Sear, M. Shenk, E.A. Smith, C. von Rueden, P. Wiessner,
{``Intergenerational Wealth Transmission and the Dynamics of Inequality in Small-Scale Societies''},
{\it Science}
{\bf 326}
(2009)
pp. 682--688.
}
\end{thebibliography}

\appendix

\section{Understanding the distribution $\Xi$ as the limit of a sequence of functions}
\label{sec:Xi}

The distribution $\Xi(w)$ was introduced in Subsection~\ref{ssec:Xi} as a linear functional.  A more complete description of it as a linear functional was given in Appendix C of an earlier paper~\cite{bib:Boghosian1}.  In this appendix we take a more pedestrian approach by describing it as the limit of a sequence of functions.

\begin{figure}
\begin{center}
\includegraphics[bbllx=0,bblly=0,bburx=225,bbury=314,width=.30\textwidth]{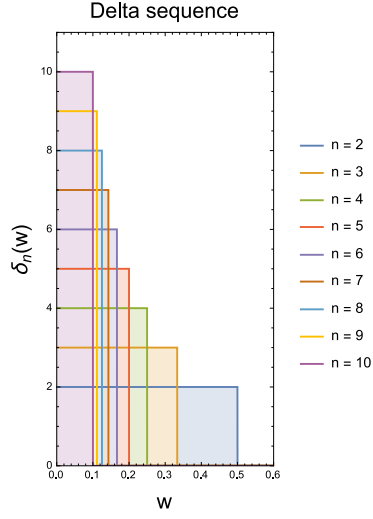}
\end{center}
\caption{{\bf Delta sequence} of step functions that are increasingly tall and narrow}
\label{fig:deltaSequence}
\end{figure}

To first remind ourselves of a more familiar example, note that we can imagine the singular $\delta$ distribution as the limit of a sequence of functions,
\begin{equation}
\delta(w) = \lim_{n\rightarrow\infty} \delta_n(w).
\end{equation}
There are many ways to define the functions $\delta_n(w)$; perhaps the simplest is
\begin{equation}
\delta_n(w) := \left\{
\begin{array}{ll}
n & \mbox{if $0<w<1/n$}\\
0 & \mbox{otherwise.}
\end{array}
\right.
\end{equation}
This sequence of functions is illustrated in Fig.~\ref{fig:deltaSequence}.  Now the zeroth moment of $\delta_n(w)$ is equal to one for all $n$, and hence equal to one in the limit as $n\rightarrow\infty$.  The first moment of $\delta_n(w)$ is easily calculated to be $1/(2n)$, and hence goes to zero in the limit as $n\rightarrow\infty$, as do all higher moments.  So we write $\int_0^\infty dw\;\delta(w) = 1$, $\int_0^\infty dw\;\delta(w) w = 0$, etc.

\begin{figure}
\begin{center}
\includegraphics[bbllx=0,bblly=0,bburx=320,bbury=204,width=.40\textwidth]{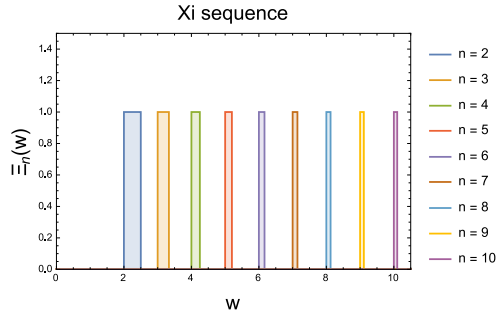}
\end{center}
\caption{{\bf Xi sequence} of square waves of unit height that are increasingly narrow and increasingly distant from the origin}
\label{fig:XiSequence}
\end{figure}

The singular distribution $\Xi(w)$ is less well known than the delta distribution, but we can also think of it as the limit of a sequence of functions.  As was the case with $\delta(w)$, there are many ways to choose this sequence, but the following one is particularly simple,
\begin{equation}
\Xi(w) = \lim_{n\rightarrow\infty} \Xi_n(w),
\end{equation}
where we have defined
\begin{equation}
\Xi_n(w) := \left\{
\begin{array}{ll}
1 & \mbox{if $n<w<n+1/n$}\\
0 & \mbox{otherwise.}
\end{array}
\right.
\end{equation}
This sequence of functions is illustrated in Fig.~\ref{fig:XiSequence}.  Now the zeroth moment of $\Xi_n(w)$ is equal to $1/n$, and hence goes to zero in the limit as $n\rightarrow\infty$.  The first moment of $\delta_n(w)$ is equal to $1+1/(2n^2)$, and hence goes to one in the limit as $n\rightarrow\infty$.  The second moment is $n+1/n+1/(3n^3)$ which diverges as $n\rightarrow\infty$, as do all higher moments.  So we write $\int_0^\infty dw\;\Xi(w) = 0$, $\int_0^\infty dw\;\Xi(w) w = 1$, etc.

\section{Numerical solution of the steady-state Fokker-Planck equation}
\label{sec:numerical}

In this section, we provide a brief description of the numerical method used to obtain steady state solutions of the Fokker-Planck equations for wealth.  We focus on Eq.~(\ref{eq:fp3CanonicalSS}), with the Pareto potentials given by Eqs.~(\ref{eq:ACanonicalSS}), (\ref{eq:LCanonicalSS}) and (\ref{eq:BCanonicalSS}), for the agent density function in the presence of redistribution and WAA, since it is the most general case.

Our first goal is to turn these four equations into a set of simultaneous ordinary differential equations, rather than an integrodifferential equation.  We do this by differentiating Eqs.~(\ref{eq:ACanonicalSS}), (\ref{eq:LCanonicalSS}) and (\ref{eq:BCanonicalSS}),
\begin{eqnarray}
\frac{dA}{dw}
&=&
-P\\
\frac{dL}{dw}
&=&
wP\\
\frac{dB}{dw}
&=&
\frac{w^2}{2}P.
\end{eqnarray}
Next, define the new dependent variable
\begin{equation}
\mu(w) := \left[B(w) + \frac{w^2}{2}A(w)\right]P(w),
\end{equation}
so that Eq.~(\ref{eq:fp3CanonicalSS}) can be combined with the above to yield the system of equations
to obtain the system of equations
\begin{eqnarray}
\frac{d\mu}{dw}
&=&
\frac{\taubar\left(1-w\right)\mu
-2\zetabar\left[\left(B-\frac{w^2}{2}A\right) + w\left(\frac{1}{2}-L\right)\right]\mu}
{B + \frac{w^2}{2}A}
\label{eq:ShootMu}\\
\frac{dA}{dw}
&=&
-\frac{\mu}{B + \frac{w^2}{2}A}\\
\frac{dL}{dw}
&=&
\frac{w\mu}{B + \frac{w^2}{2}A}\\
\frac{dB}{dw}
&=&
\frac{\frac{w^2}{2}\mu}{B + \frac{w^2}{2}A}.
\label{eq:ShootB}
\end{eqnarray}
These constitute four simultaneous ordinary differential equations in four unknowns.  The initial conditions are
\begin{eqnarray*}
\mu(0) &=& 0\\
A(0) &=& 1\\
L(0) &=& 0\\
B(0) &=& 0,
\end{eqnarray*}
and we expect the final condition
\begin{equation}
\lim_{w\rightarrow\infty} A(w) = 0.
\label{eq:AFinal}
\end{equation}
The above final condition assures the normalization $N = \int_0^\infty dw\; P(w) = 1$.  We may also expect that
\begin{equation}
\lim_{w\rightarrow\infty} L(w) = 1,
\label{eq:LFinal}
\end{equation}
but, for reasons described below, this final condition is not as fundamental as Eq.~(\ref{eq:AFinal}).

Eqs.~(\ref{eq:ShootMu}) through (\ref{eq:ShootB}), supplemented with Eq.~(\ref{eq:AFinal}), may be regarded as a boundary-value problem.  We would like to solve this using a shooting method.  Unfortunately, there is a problem with re-interpreting these equations as an initial-value problem.  As noted in the analysis of the steady-state Fokker-Planck equation for small $w$, the solution is non-analytic at the origin.  As a consequence, it is non-unique at the origin.  It is clear, for example, that $\mu(w)=0$ will solve Eq.~(\ref{eq:ShootMu}).  This is not the solution we want, since it would imply that $P=0$, and hence it would not satisfy the final condition, Eq.~(\ref{eq:AFinal}).

To surmount this difficulty, we use the asymptotic solution, Eq.~(\ref{eq:fp1ssLowWWAA}), in the vicinity of the origin, say for $w\in[0,\delta w]$, and then use the final conditions in this interval as initial conditions to Eqs.~(\ref{eq:ShootMu}) through (\ref{eq:ShootB}).  Reflective of the above-described non-uniqueness, the asymptotic solution, Eq.~(\ref{eq:fp1ssLowWWAA}), has an arbitrary constant $C_0$ that is not fixed by the initial condition.  The strategy is then to vary $C_0$ until the final condition, Eq.~(\ref{eq:AFinal}), is satisfied.  In practice, this can be done by placing the ordinary differential equation solver inside the loop of a Newton-Raphson solver for $C_0$.

An initial guess for $C_0$ may be found by examining a graph of $\lim_{w\rightarrow\infty} A(w)$ versus $C_0$ and looking for zero crossings.  In practice, this must be done by hand.  The initial guess must be rather accurate for the method to converge, and we have not found a way to automate its determination.

When the method does converge, however, it is remarkably accurate.  One way of checking its accuracy is to examine $\lim_{w\rightarrow\infty} L(w)$.  We found that Eq.~(\ref{eq:LFinal}) was accurate to within eight significant digits, but only for $\zetabar < \taubar$.  For general $\zetabar$, we found that this limit was given by Eq.~(\ref{eq:LFinalGen}), also to within eight significant digits.  In addition to giving us confidence that our solutions were accurate, the transition observed in Eq.~(\ref{eq:LFinalGen}) was how we first discovered the phenomenon of criticality described in this paper.

Finally, we note that the domain $w\in[0,\infty)$ is non-compact and hence inconvenient to use numerically for the above-described calculations.  For this reason, we first transform the independent variable from $w$ to $\theta=w/(1+w)$, and we consider $\theta\in[0,1]$, with final conditions imposed at $\theta=1$ in all of the above calculations.

\section{Asymptotic analysis of steady state at large $w$}
\label{sec:LargeW}
\subsubsection{Analysis for subcritical WAA}

The asymptotic analysis of Eq.~(\ref{eq:fp3CanonicalSS}) for very large $w$ is significantly more difficult than that for very small $w$.  To begin, we rearrange that equation as follows,
\begin{equation}
\frac{d}{dw}\ln P
=
\frac{
\taubar\left(1-w\right)
-2\zetabar\left[\left(B-\frac{w^2}{2}A\right) + w\left(\frac{1}{2}-L\right)\right]
-wA}
{B + \frac{w^2}{2}A}.
\label{eq:fp3CanonicalSS2}
\end{equation}

For $\zetabar < \taubar$, our approach shall be to posit a gaussian distribution of the form
\begin{equation}
P(w) \approx C_\infty\exp\left(-a w^2 - b w\right),
\label{eq:PPosited}
\end{equation}
valid for very large $w$ only, where $a>0$ and $b$ are constants to be determined.  We must justify this form a posteriori.  Substituting it on the left-hand side of Eq.~(\ref{eq:fp3CanonicalSS2}) immediately yields $-2aw-b$.  We must show that substituting it on the right-hand side of that equation also yields a linear function of $w$, plus subdominant corrections.  By equating coefficients of the dominant linear terms, we will be able to determine $a$ and $b$.

To substitute Eq.~(\ref{eq:PPosited}) on the right-hand side of Eq.~(\ref{eq:fp3CanonicalSS2}) requires some preliminary work.  We must first calculate $A$, $L$ and $B$ for very large $w$.  Since the first of these is defined by an integral from $w$ to infinity, we can substitute Eq.~(\ref{eq:PPosited}) directly into Eq.~(\ref{eq:ACanonicalSS}) to obtain
\begin{equation}
A(w) \approx \frac{C_\infty}{2}\sqrt{\frac{\pi}{a}}\exp\left(\frac{b^2}{4a}\right)
\mbox{erfc}\left(\frac{2a w+b}{2\sqrt{\alpha}}\right),
\label{eq:APosited}
\end{equation}
where $\mbox{erfc}$ denotes the complementary error function.

Because $\lim_{w\rightarrow\infty}L(w)=1$, we may next rewrite Eq.~(\ref{eq:LCanonicalSS}) as
\begin{equation}
L(w) = 1 - \int_w^\infty dx\; P(x) x,
\label{eq:LComplement}
\end{equation}
and substitute Eq.~(\ref{eq:PPosited}) in the above to obtain
\begin{equation}
L(w) \approx 1-\frac{C_\infty}{2a}\exp\left(-a w^2 - b w\right)
+\frac{C_\infty b}{4a}\sqrt{\frac{\pi}{a}}\exp\left(\frac{b^2}{4a}\right)
\mbox{erfc}\left(\frac{2a w+b}{2\sqrt{\alpha}}\right).
\label{eq:LPosited}
\end{equation}

Next note that a consequence of Eq.~(\ref{eq:PPosited}) is that $B_\infty := \lim_{w\rightarrow\infty}B(w) > 0$ is finite, so we may rewrite Eq.~(\ref{eq:BCanonicalSS}) as follows
\begin{equation}
B(w) = B_\infty - \int_w^\infty dx\; P(x)\frac{x^2}{2}.
\end{equation}
Inserting Eq.~(\ref{eq:PPosited}) results in
\begin{equation}
B(w) \approx B_\infty +\frac{C_\infty}{8a^2}(b-2aw)\exp\left(-a w^2 - b w\right)
-\frac{C_\infty}{16a^2}(b^2+2a)\sqrt{\frac{\pi}{a}}\exp\left(\frac{b^2}{4a}\right)
\mbox{erfc}\left(\frac{2a w+b}{2\sqrt{\alpha}}\right)
\label{eq:BPosited}
\end{equation}

We are now ready to substitute Eqs.~(\ref{eq:APosited}), (\ref{eq:LPosited}) and (\ref{eq:BPosited}) into the right-hand side of Eq.~(\ref{eq:fp3CanonicalSS2}).  The resulting expression may be simplified by using the asymptotic expansion for the complementary error function,
\begin{equation}
\mbox{erfc}(z)\sim\frac{\exp(-z^2)}{\sqrt{\pi}\; z}.
\label{eq:erfcAsymp}
\end{equation}
After a considerable amount of algebra, the right-hand side of Eq.~(\ref{eq:fp3CanonicalSS2}) reduces to
\begin{equation}
-\left(\frac{\taubar-\zetabar}{B_\infty}\right)w
-\left(\frac{2B_\infty\zetabar-\taubar}{B_\infty}\right)
+\calO\left(e^{-aw^2}\right),
\end{equation}
which demonstrates that the dominant part of the right-hand side of Eq.~(\ref{eq:fp3CanonicalSS2}) is indeed simply a linear function of $w$.  Demanding that this equal $-2aw-b$ for all $w$ immediately yields
\begin{eqnarray}
a &=& \frac{\taubar-\zetabar}{2B_\infty}\\
b &=& \frac{2B_\infty\zetabar-\taubar}{B_\infty}.
\end{eqnarray}
As long as $\zetabar < \taubar$, we have $a>0$, providing the desired a posteriori justification of Eq.~(\ref{eq:PPosited}).

\subsubsection{Analysis for supercritical WAA}

As noted above, the preceding analysis is valid only below criticality; that is, for $\zetabar < \taubar$.  For higher values of $\zetabar$, our solution for $a$ becomes negative, invalidating the a posteriori justification of Eq.~(\ref{eq:PPosited}).

For $\zetabar > \taubar$, we posit the oligarchical solution
\begin{equation}
P(w) \approx p(w) + \left(1-\frac{\taubar}{\zetabar}\right) \Xi(w),
\label{eq:PPositedOligarchical}
\end{equation}
where
\begin{equation}
p(w) = C_\infty\exp\left(-a w^2 - b w\right),
\end{equation}
and where $1-\taubar/\zetabar$ is the fraction of the total wealth taken by the oligarch.  Again, we seek an a posteriori justification of this form.

The presence of the oligarchical term in Eq.~(\ref{eq:PPositedOligarchical}) will not change the fact that
\begin{equation}
\lim_{w\rightarrow\infty}A(w)=0,
\end{equation}
because the zeroth moment of $\Xi$ vanishes.  Hence, the asymptotic form for $A(w)$ will be unchanged.

The first moment will be altered as a result of the oligarchical term, as can be seen from the following argument,
\begin{eqnarray}
L(w)
&=&
1-\int_w^\infty dx\;P(x) x\\
&=&
1-\int_w^\infty dx\;\left[p(x) + \left(1-\frac{\taubar}{\zetabar}\right) \Xi(x)\right]x\\
&=&
1-\int_w^\infty dx\;p(x)x
-\left(1-\frac{\taubar}{\zetabar}\right)\int_w^\infty dx\;\Xi(x)x\\
&=&
1-\int_w^\infty dx\;p(x)x
-\left(1-\frac{\taubar}{\zetabar}\right)\\
&=&
\frac{\taubar}{\zetabar}-\int_w^\infty dx\;p(x)x.
\end{eqnarray}
In other words, the supercritical value of $L(w)$ is $\frac{\taubar}{\zetabar}-1$ plus the subcritical version, or
\begin{equation}
L(w)\approx
\frac{\taubar}{\zetabar}-\frac{C_\infty}{2a}\exp\left(-a w^2 - b w\right)
+\frac{C_\infty b}{4a}\sqrt{\frac{\pi}{a}}\exp\left(\frac{b^2}{4a}\right)
\mbox{erfc}\left(\frac{2a w+b}{2\sqrt{\alpha}}\right).
\label{eq:LPositedSup}
\end{equation}

Next, we turn our attention to $B(w)$.  The second moment of $\Xi$ does not exist, but this is not a problem.  For any {\it finite} $w$, Eq.~(\ref{eq:BCanonicalSS}) yields
\begin{equation}
B(w) = \int_0^w dx\; p(x)\frac{x^2}{2},
\end{equation}
which we may rewrite as
\begin{equation}
B(w) = B_\infty - \int_w^\infty dx\; p(x)\frac{x^2}{2}.
\end{equation}
Consequently, Eq.~(\ref{eq:BPosited}) is unchanged.

So, our posited forms for $A(w)$ and $B(w)$ are still given by Eqs.~(\ref{eq:APosited}) and (\ref{eq:BPosited}), whereas $L(w)$ is now given by Eq.~(\ref{eq:LPositedSup}).  Substituting these expressions into the right-hand side of Eq.~(\ref{eq:fp3CanonicalSS2}), and using the asymptotic expansion for the complementary error function, Eq.~(\ref{eq:erfcAsymp}), we eventually find that the right-hand side of Eq.~(\ref{eq:fp3CanonicalSS2}) reduces to
\begin{equation}
-\left(\frac{\zetabar-\taubar}{B_\infty}\right)w
-\left(\frac{2B_\infty\zetabar-\taubar}{B_\infty}\right)
+\calO\left(e^{-aw^2}\right).
\end{equation}
It follows that
\begin{eqnarray}
a &=& \frac{\zetabar-\taubar}{2B_\infty}\\
b &=& \frac{2B_\infty\zetabar-\taubar}{B_\infty}.
\end{eqnarray}
Note that the value of $a$ is once again positive, providing the a posteriori justification of this form.  Interestingly, $b$ is unchanged from the subcritical case.

In the end, both the subcritical and supercritical results can be combined into the single expression
\begin{eqnarray}
a &=& \frac{\left|\taubar-\zetabar\right|}{2B_\infty}\\
b &=& \frac{2B_\infty\zetabar-\taubar}{B_\infty},
\end{eqnarray}
so a plot of $a$ versus $\zetabar$ would exhibit a slope discontinuity at the critical point $\zetabar = \taubar$.  We may also conclude that the tail of the distribution decays as a gaussian both above and below criticality, but precisely at criticality it decays exponentially.

\end{document}